\documentclass[aps,twocolumn,preprintnumbers,longbibliography]{revtex4-1}
\usepackage{amsmath,amssymb}
\usepackage{bm}
\usepackage{comment}
\usepackage[svgnames]{xcolor}
\usepackage[colorlinks]{hyperref}
\hypersetup{linkcolor=DarkBlue,citecolor=DarkBlue,filecolor=black,urlcolor=DarkBlue}
\newcommand\x{\mathbf{x}}
\newcommand\y{\mathbf{y}}
\newcommand\p{\mathbf{p}}
\newcommand\q{\mathbf{q}}
\renewcommand\d{\partial}
\newcommand\+{\dagger}

\newcommand\C{\mathcal{C}}
\renewcommand\P{\mathcal{P}}
\newcommand\T{\mathcal{T}}
\newcommand\<{\langle}
\renewcommand\>{\rangle}

\begin{document}

\preprint{EFI 15-6}
\title{Is the Composite Fermion a Dirac Particle?}

\author{Dam Thanh Son}
\affiliation{Kadanoff Center for Theoretical Physics, University of Chicago, Chicago, Illinois 60637, USA}
\date{February 2015, revised July 2015}

\begin{abstract}
  We propose a particle-hole symmetric theory of the Fermi-liquid
  ground state of a half-filled Landau level.  This theory should be
  applicable for a Dirac fermion in the magnetic field at charge
  neutrality, as well as for the $\nu=\frac12$ quantum Hall ground
  state of nonrelativistic fermions in the limit of negligible
  inter-Landau-level mixing.  We argue that when particle-hole
  symmetry is exact, the composite fermion is a massless Dirac
  fermion, characterized by a Berry phase of $\pi$ around the Fermi
  circle.  We write down a tentative effective field theory of such a
  fermion and discuss the discrete symmetries, in particular,
  $\C\P$. The Dirac composite fermions interact through a gauge, but
  non-Chern-Simons, interaction.  The particle-hole conjugate pair of
  Jain-sequence states at filling factors $\frac n{2n+1}$ and
  $\frac{n+1}{2n+1}$, which in the conventional composite fermion
  picture corresponds to integer quantum Hall states with different
  filling factors, $n$ and $n+1$, is now mapped to the same
  half-integer filling factor $n+\frac12$ of the Dirac composite
  fermion.  The Pfaffian and anti-Pfaffian states are interpreted as
  $d$-wave Bardeen-Cooper-Schrieffer paired states of the Dirac
  fermion with orbital angular momentum of opposite signs, while
  $s$-wave pairing would give rise to a particle-hole symmetric
  non-Abelian gapped phase.  When particle-hole symmetry is not exact,
  the Dirac fermion has a $\C\P$-breaking mass.  The conventional
  fermionic Chern-Simons theory is shown to emerge in the
  nonrelativistic limit of the massive theory.
\end{abstract}

\maketitle

\section{Introduction}

The theory of the fractional quantum Hall (FQH)
effect~\cite{Tsui:1982yy,Laughlin:1983fy} is based on the paradigm of
the composite
fermion~\cite{Jain:1989tx,Fradkin:1991wy,Halperin:1992mh}, which
provides a unified explanation of a large amount of observed
phenomena, among which the most early ones are the Jain
sequences---series of quantum Hall plateaux at filling factors $\nu$
near $1/2$, $1/4$ etc.  The composite fermion picture gives rise
to extremely accurate wave functions of FQH ground states~\cite{Jain-book}.

At half filling, the composite fermion (CF)
picture, developed into a mathematical framework of the Chern-Simons (CS) field theory
by Halperin, Lee, and Read (HLR)~\cite{Halperin:1992mh},
predicts that the ground state is a Fermi liquid, providing an
explanation for the results of acoustic-wave-propagation
experiments~\cite{Willett:1990}.  Near, but not exactly at half
filling, the CF is predicted to feel a small residual magnetic field,
and the semiclassical motion of the CF in such a field has been
observed experimentally~\cite{Kang:1993,Goldman:1994zz}.  The
composite fermion theory also provides an elegant interpretation of
the Pfaffian (or Moore-Read) state~\cite{Moore:1991ks} as a $p_x+ip_y$
Bardeen-Cooper-Schrieffer (BCS) paired state~\cite{Read:1999fn}.

Despite its success, one of the symmetries of FQH systems in the limit of
zero Landau-level (LL) mixing---the particle-hole
symmetry~\cite{Girvin:1984zz}---is not explicit within the CF paradigm.  The CF
is constructed by attaching a magnetic flux to the electron before
projecting to the lowest Landau level (LLL); the
CS field theory formalism, strictly speaking, does
not allow one to attach fluxes to holes in the LLL.  In one
manifestation of the particle-hole asymmetric nature of the formalism,
the two Jain-sequence states with $\nu=\frac n{2n+1}$ and
$\nu=\frac{n+1}{2n+1}$, which form a particle-hole conjugate pair,
receive slightly different interpretations in the CF language: the
former fraction is an integer quantum Hall (IQH) state of CFs with $n$
filled Landau levels, while in the latter $n+1$ Landau levels are
filled.  A related issue of the CF picture is its failure to account
for the anti-Pfaffian state~\cite{Levin:2007,SSLee:2007}---the
particle-hole conjugate of the Moore-Read state---in a simple manner.

The zero bare mass limit, where particle-hole symmetry is exact, is
particularly difficult to analyze within the HLR theory.  Kivelson \emph{et
al.}~\cite{Kivelson:1997} analyzed the HLR theory at $\nu=\frac12$ and
found that particle-hole symmetry requires the liquid of CFs to have a
Hall conductivity $\sigma_{xy}^{\rm CF}=-\frac12$.  In a zero net
magnetic field, this means that the CF liquid has an anomalous Hall
coefficient and seems to contradict the Fermi liquid nature of the CFs.
Kivelson \emph{et al.}\ did not find any set of Feynman diagrams that could
lead to a nonzero $\sigma_{xy}^{\rm CF}$.  As one possible solution,
they proposed that the problem lies in the noncommutativity of the
limit of $m\to0$ (the LLL limit) and the limit of taking the density
of impurities to zero (the clean limit).   This proposal leaves
unanswered the question of why $\sigma_{xy}^{\rm CF}$ is exactly
$-\frac12$ at all frequencies, independent of the physics of
impurities.  Few attempts have been made to resolve the apparent
inability to fit particle-hole symmetry to the CF picture; examples
include Ref.~\cite{Lee:1998zze}.

One logical possibility is that particle-hole symmetry is
spontaneously broken and the HLR theory describes only one of the two
$\nu=\frac12$ ground states, which become degenerate at zero Landau
level mixing.  In this case, the anti-Pfaffian state would be
inaccessible from within the HLR theory.  Numerical simulations,
however, seem to be consistent with a particle-hole symmetric
$\nu=\frac12$ ground state~\cite{Rezayi:2000zz}.
Away from half filling, despite the apparent asymmetry in the treatments of the
$\nu=\frac n{2n+1}$ and $\nu=\frac{n+1}{2n+1}$ states, particle-hole
symmetry maps one CF trial wave function to another with high
accuracy~\cite{Wu:1993}.  In addition, a recent experiment,
set up to measure the Fermi momentum using commensurability effects in a
periodic potential, implies that the Fermi momentum is determined by
the density of particles at $\nu<\frac12$ and of holes at
$\nu>\frac12$~\cite{Baldwin:2014}.  One interpretation of this
experiment is that the $\nu=\frac12$ state allows two alternative, but
equivalent, descriptions as a Fermi liquid of either particles or holes.
This would mean that the $\nu=\frac12$ ground state coincides with its
particle-hole conjugate.

The problem of particle-hole symmetry is more acute for systems with
Dirac fermions.  The Dirac fermion is realized in graphene and on the
surface of topological insulators (TIs) and is a relatively new venue
for studying the quantum Hall (QH) effect.  For Dirac fermions,  the IQH
plateaux occur at half-integer values of the Hall conductivity:
$\sigma_{xy}=(n+\frac12)\frac{e^2}h$, which (after accounting for the
fourfold degeneracy of the Dirac fermion) has been seen in
graphene~\cite{Novoselov:2005kj,Zhang:2005zz}.  FQH plateaux have also
been observed in graphene~\cite{FQHE-graphene1,FQHE-graphene2}; there
is intriguing evidence that FQH effect may exist on the surface of
TIs~\cite{Boebinger:2010}.  In contrast to the nonrelativistic case,
for Dirac fermions, particle-hole symmetry is a good symmetry even with
Landau level mixing.  It is not obvious that the flux attachment
procedure can be carried out for Dirac fermions; the usual
workaround is to work in the limit of zero Landau-level mixing where
the projected Hamiltonian is identical to the nonrelativistic
one~\cite{Khveshchenko:2006}.  This method explicitly breaks
particle-hole symmetry and, furthermore, does not work at finite Landau
level mixing.

In this paper we propose
an explicitly particle-hole symmetric effective theory describing
the low-energy dynamics of the Fermi liquid state of a half-filled Landau level.  This theory is constructed to
respect all discrete symmetries and to satisfy phenomenological
constraints, in particular, the existence of the two Jain sequences
below and above $\nu=\frac12$.  Our proposal is similar to the
fermionic Chern-Simons (HLR) theory in that the elementary degree of
freedom is a fermion; however, it differs from it in several ways:
\begin{itemize}
\item[(i)] The fermion is, by nature, a Dirac fermion.
\item[(ii)] The fermion is its own particle-hole conjugate.
  \item[(iii)] The fermion has no Chern-Simons interactions.  This is
    an important point as the Chern-Simons term is not consistent
    with particle-hole symmetry~\footnote{Note that in Read's LLL theory
      of the bosonic $\nu=1$ state~\cite{Read:1998dn} there is also a gauge field with
      no Chern-Simons term.}.
\end{itemize}
As per point (i), one may wonder if there is any difference between the
Fermi liquid of Dirac fermions and that of nonrelativistic fermions:
both are characterized by a linear dispersion relation of
quasiparticles near the Fermi surface.  The difference is in the Berry
phase when the quasiparticle is moved around the Fermi surface (a
circle in 2D), which is $\pm\pi$ in the case of a Dirac fermion.  The
importance of the Berry phase as a property of the Landau fermion
quasiparticle was emphasized by Haldane~\cite{Haldane:2004zz}.

The Berry phase of $\pi$ offers a resolution to the puzzle of the
anomalous Hall conductivity of the CFs.  As shown by
Haldane~\cite{Haldane:2004zz}, the unquantized part of the anomalous
Hall conductivity of the CF Fermi liquid is equal to the global Berry
phase $\gamma$ which quasiparticle receives when it moves around the Fermi
disk,
\begin{equation}
  \sigma_{xy}^{\rm CF} = \left( \frac\gamma{2\pi} + n\right) \frac{e^2}h \,,
  \quad n\in\mathbb{Z}.
\end{equation}
When $\gamma=\pm\pi$, this equation is consistent with
$\sigma_{xy}^{\rm CF}=-\frac12$.  The resolution to the puzzle  is not the noncommutativity of the $m\to0$
limit and the clean limit; rather, it is in an ingredient of the Fermi
liquid theory missing in all treatments of the CFs so far: the global
Berry phase.

Beside these differences, there are also many similarities between the picture proposed here and
the standard CF picture.  In particular, the fermion quasiparticle is
electrically neutral.  The Jain-sequence states near half filling
are mapped to IQH states of the
CFs; however, in our theory, particle-hole conjugate states map to
IQH states with the same half-integer filling factors.


We show that the theory
suggests the existence of a
particle-hole symmetric gapped state at $\nu=\frac12$, distinct from
the Pfaffian and anti-Pfaffian states. 

The structure of the paper is as follows.  In Sec.~\ref{sec:model})
we start by writing down a model consisting of a
single two-component Dirac fermion, localized on a (2+1)D ``brane,''
coupled to electromagnetism in four dimensions.  This model does not
contain complications specific for graphene or the surface state of
TIs (the fourfold degeneracy in graphene, or the Zeeman coupling in
the case of TIs) and should realize a Fermi-liquid state in a finite
magnetic field.  We show that the electromagnetic response in
this model is related directly to that of the nonrelativistic
electrons on the LLL.
We use the model to discuss discrete symmetries, which are
also shared by the nonrelativistic model in the LLL limit,
emphasizing
that there are two independent discrete symmetries in the finite magnetic field
at charge neutrality: $\C\P$ and $\P\T$.
We then put forward, in Sec.~\ref{sec:proposal}, our proposal
for the low-energy effective field theory and discuss physical implications.
 The possible connection to the
conventional fermionic Chern-Simons theory is discussed in
Sec.~\ref{sec:comp}.  Finally, Sec.~\ref{sec:concl} contains
concluding remarks.

\section{A relativistic model realizing the $\nu=\frac12$ FQH state.}
\label{sec:model}

To be more specific, we discuss a theory of a massless fermion
localized on a (2+1)-dimensional brane placed at $z=0$,
interacting through a U(1) gauge field in the (3+1)-dimensional bulk.
\begin{equation}\label{S}
  S = \int\!d^3x\, i\bar\Psi\gamma^\mu(\d_\mu-iA_\mu)\Psi
      - \frac1{4e^2}\int\!d^4x\, F_{\mu\nu}^2\,,
\end{equation}
where $\Psi$ is a two-component spinor.  We choose the following
representation for the gamma matrices:
\begin{equation}
  \gamma^0 = \sigma^3, \quad \gamma^1 = i\sigma^2, \quad
  \gamma^2 = -i\sigma^1.
\end{equation}
One can think of $\Psi$ as the fermion zero mode localized on a
domain wall.  In the condensed-matter language, $\Psi$ is the surface
mode of a 3D TI.  The theory has one dimensionless coupling constant
$e$.  In contrast to the usual QED, $e$ does not run since it
determines the strength of the electromagnetic interactions infinitely
far away from the brane, where there is no matter that would
renormalize it.  We are mostly interested in the weak coupling
regime where $e^2\ll1$, but most of our statements should be valid up to
some finite value of $e^2$.

Our task is to understand the ground state of the system in the finite
magnetic field $B=F_{xy}$ and its excitations.  This problem is
nonperturbative even at $e^2\ll1$ since it maps to a FQH problem.  To
see this, assume at first that $e^2=0$ and recall that the Landau levels
of the Dirac Hamiltonian are $E = \pm \sqrt{2nB}$.  In the ground
state, the states with negative energy are filled and those with
positive energy are empty, but the noninteracting Hamiltonian gives
us no prescription for the $n=0$ LL (the zeroth LL), whose states have
exact zero energy.  The true ground state is only singled out when
interaction is turned on.  It is worth noting that the energy scale
induced by the interaction is $e^2\sqrt B$, while the spacing between LLs
is $\sqrt B$.

To see that the problem maps to the $\nu=\frac12$ QH problem, we
notice that, because of particle-hole symmetry, at zero chemical potential,
the zeroth LL must be half full (or half empty).  At small $e^2$, all
essential physics occurs on the zeroth LL, and the Hamiltonian
projected to this LL is the same as the usual projected QH
Hamiltonian since the Dirac orbitals on the zeroth LL are the same
(neglecting one component of the Dirac spinor which vanishes for these
states) as the Landau orbitals of a nonrelativistic particle [see
  Eq.~(\ref{orbital}) below].

\subsection{Equivalence between Dirac fermions and nonrelativistic
  fermions on the LLL}

\subsubsection{Universality of the LLL limit}

We have argued that the problem of finding the ground state of a
Dirac fermion in a magnetic field, in the weak coupling regime
$e^2\ll1$, is equivalent to finding the ground state of a nonrelativistic
fermion in the LLL limit $m\to0$.  In this subsection, we show that
the equivalence can be extended further: the full electromagnetic
response of one theory can be obtained from that of another theory.
It is obvious that the density response to an external potential is
the same in the two theories.  The relationship between the currents in
the two theories is, however, slightly more complicated.

To establish the connection, we recall a recent procedure to
derive the expression for the current in the $m\to0$ limit of a
nonrelativistic theory~\cite{Nguyen:2014uea}.  We start from the
Lagrangian
\begin{equation}\label{Lg=0}
  \mathcal L  = i \psi^\+ D_t \psi - \frac1{2m} D_i\psi^\+ D_i\psi
  + \mathcal L_{\rm int}.
\end{equation}
In later formulas, we drop the interaction term $\mathcal L_{\rm
  int}$, always implicitly implying it. 
We now add to the Lagrangian a magnetic-moment term, giving
the particle a gyromagnetic factor $g=2$ (and assuming the system is
fully spin polarized),
\begin{equation}\label{Lg=2}
  \mathcal L_{g=2}  = i \psi^\+ D_t \psi - \frac1{2m} D_i\psi^\+ D_i\psi
  + \frac B{2m} \psi^\+\psi \,.
\end{equation}
For a constant magnetic field, the added term is proportional to the
total number of particles, which commutes with the Hamiltonian and does
not alter the dynamics.  However, the expression for the
electromagnetic current $j_i=\delta S/\delta A_i$ has changed.
Denoting the current corresponding to Eq.~(\ref{Lg=0}) by $j^i_{g=0}$
and to Eq.~(\ref{Lg=2}) by $j^i_{g=2}$, we find
\begin{equation}
  j^i_{g=0} = j^i_{g=2} - \frac1{2m}\epsilon^{ij}\d_j\rho
  \,.
\end{equation}
The two currents differ only by a solenoidal term, and if we know
the density response, we can get one current from the other.  As we
will see later, $j^i_{g=2}$ is finite in the limit $m\to0$, but $j^i_{g=0}$
is, in general, divergent.  In the subsequent discussion, by the
electromagnetic current we have in mind $j^i_{g=2}$ by default.
Note that for linear response at $q=0$ (but any frequency), the two
currents coincide.

Ignoring a total derivative, the Lagrangian~(\ref{Lg=2}) can be
rewritten as
\begin{equation}
  \mathcal L = i\psi^\+ D_t\psi - \frac1{2m} (D_x-iD_y) \psi^\+ (D_x+iD_y)\psi
\,.
\end{equation}
Introducing auxiliary fields $\chi$ and $\chi^\+$, we can recast the
Lagrangian in the form
\begin{align}
  \mathcal L = & i\psi^\+ D_t\psi + i\psi^\+(D_x-iD_y)\chi
      + i\chi^\+(D_x+iD_y)\psi\nonumber\\
      & + 2m\chi^\+ \chi\,.
\end{align}
Now, taking the $m\to0$ limit, the Lagrangian becomes
\begin{equation}\label{L-LLL}
  \mathcal L =  i\psi^\+ D_t\psi + i\psi^\+(D_x-iD_y)\chi
      + i\chi^\+(D_x+iD_y)\psi\,.
\end{equation}

The last form of the Lagrangian makes it clear that the limit $m\to0$
is finite.  In fact, Eq.~(\ref{L-LLL}) can be thought of as a
Lagrangian formulation of the problem on the lowest Landau level.  The
variables $\chi$ and $\chi^\+$ are Lagrange multipliers enforcing the
constraints
\begin{equation}
  (D_x+iD_y)\psi = 0, \quad (D_x-iD_y)\psi^\+ = 0,
\end{equation}
which is nothing but the LLL constraint.  For example, in the
symmetric gauge $A_x=-\frac12 By$, $A_y=\frac12 Bx$, the constraints
imply that $\psi$ is a linear combination of Landau's orbitals $z^n
e^{-B|z|^2/4}$, $n=0,1,2\ldots$  Equation~(\ref{L-LLL}) implies that the
physics on the LLL is universal, i.e., independent of how the LLL
limit is approached.  In Ref.~\cite{Nguyen:2014uea}, it was shown
that the current computed from~Eq.~(\ref{L-LLL}) coincides with that found
in Ref.~\cite{Martinez:1993xv}.  This current differs from the
one obtained in Ref.~\cite{RajaramanSondhi} by a solenoidal
term, which is not surprising since Ref.~\cite{RajaramanSondhi}
assumes a zero geomagnetic factor and should gives $j^i_{g=0}$ instead
of $j^i_{g=2}.$

Let us now consider the Dirac field theory.  Denoting the two components
of the Dirac spinor as
\begin{equation}
  \psi = \begin{pmatrix}  \psi \\ \chi \end{pmatrix} ,
\end{equation}
the Dirac action can be written as
\begin{multline}
  \mathcal L = i\psi^\+ D_t \psi
  + i \chi^\+ D_t \chi \\ + i\psi^\+(D_x-iD_y)\chi 
  + i\chi^\+(D_x+iD_y)\psi .
\end{multline}
If one concentrates on states on the zeroth Landau level, the Dirac
spinor will have $\chi$ much smaller than $\psi$.  In the Lagrangian,
the term $\chi^\+\d_t\chi$ is quadratic in $\chi$ and can be
neglected compared to other terms.  Terms linear in $\chi$ will still
have to be kept since the time derivative $D_t$ is of the same order of
smallness as $\chi$.  Thus, the Lagrangian becomes exactly
Eq.~(\ref{L-LLL}).

\subsubsection{Relationship between currents in relativistic and
  nonrelativistic theories}

Although the expressions of the current coming from the modes on the
LLL in the relativistic and nonrelativistic theories are the same, the
physics of the relativistic model is slightly different from that of
the nonrelativistic one by the presence of a Dirac sea of Landau
levels with negative energies.  The states in the Dirac sea
contribute a uniform charge density $-B/4\pi$; thus,
\begin{equation}\label{rDrNR}
  \rho_{\rm D} = \rho_{\rm NR} - \frac B{4\pi} \,,
\end{equation}
where the subscripts ``D'' and ``NR'' refer to ``Dirac'' and
``nonrelativistic,'' respectively.  To find the contribution of the
sea to the current, we notice that the electrons in the Dirac sea are
subject to an additional scalar potential created by the electrons
on the zeroth
Landau level,
\begin{equation}\label{V-Hartree}
  V_{\rm eff}(\x) = \int\!d\y\, V(\x-\y) \rho(\x) \equiv (V\cdot\rho)(\x),
\end{equation}
where $V(\x-\y)$ is the interaction potential between electrons (i.e.,
the Coulomb potential).
The response of the Dirac sea to this scalar potential gives an extra
contribution to the current, leading to the following relationship
between the currents:
\begin{equation}\label{jDjNR}
  j^i_{\rm D} = j^i_{\rm NR} + \frac1{4\pi} \epsilon^{ij}\d_j (V\cdot\rho).
\end{equation}

The two formulas can be summarized as the following relationship
between the logarithms of the partition functions ($W=-i\ln Z$) in the
two theories, valid for any external $A_0$ and to linear order in
perturbations of $A_i$ around the background,
\begin{equation}
  W_{\rm D}[A_0,A_i] = -\frac{A_0 B}{4\pi} + W_{\rm NR}[A_0+\frac{V\cdot B}
    {4\pi}, A_i].
\end{equation}
Differentiating with respect to $A_0$ and $A_i$ we recover
Eqs.~(\ref{rDrNR}) and (\ref{jDjNR}).

The formulas are written in the long-wavelength limit of external
probes.  This is sufficient for the purpose of the rest of the paper.
For completeness, we mention that there are corrections to the
formulas from two sources: i) the wavenumber dependence of the Hall
conductivity of the filled levels; and ii) the exchange interactions
between the electrons in the Dirac sea and those on the zeroth Landau level
[only the Hartree interaction has been taken into account in
Eq.~(\ref{V-Hartree})].  Both effects are small in the long-wavelength
limit for the potential $V$ sufficiently long ranged so that its Fourier
transform $V(q)$ diverges at $q\to0$ (as the Coulomb potential).

\subsubsection{Relativistic convention}

The discussion above shows that there is a direct relationship between
physical observables in the cases of Dirac and nonrelativistic
fermions in the LLL limit.  Here, we collect formulas that relate
quantities in the two theories, which will be useful in our further
discussion.

We define the filling factor as the ratio
\begin{equation}
  \nu = \frac \rho{B/2\pi}\,,
\end{equation}
where $\rho$ is the charge density.  According to Eq.~(\ref{rDrNR}),
the relationship between the relativistic and nonrelativistic filling
factors is
\begin{equation}
  \nu = \nu_{\rm NR} - \frac12\,,
\end{equation}
where we have dropped the index ``D'' on the left-hand side.  In
the relativistic theory, $\nu=0$ occurs at zero chemical potential (the charge
neutrality point), where the $n=0$ Landau level is half filled.  The
latter corresponds to $\nu_{\rm NR}=1/2$.

Similarly, one can derive a relationship between the conductivities
in the relativistic and nonrelativistic theories from
Eq.~(\ref{jDjNR}).  At zero wave number (but any frequency), the
spatial derivative term on the right-hand side of Eq.~(\ref{jDjNR})
vanishes, giving
\begin{align}
  \sigma_{xx}(\omega) &= \sigma_{xx}^{\rm NR}(\omega),\\
  \sigma_{xy}(\omega) &= \sigma_{xy}^{\rm NR}(\omega) - \frac12
  \,. \label{sxy-sNRxy}
\end{align}
We have measured the conductivities in units of $e^2/h$.  One can
interpret Eq.~(\ref{sxy-sNRxy}) as the statement that the Hall
conductivity of the Dirac fermion consists of a contribution from the
zeroth Landau level, equal to $\sigma_{xy}^{\rm NR}$, and the
contribution from the filled negative-energy Landau levels, equal to
$-\frac12$.  Note that these formulas are valid also in the presence
of impurities.

Next we consider the shift.  In the relativistic context, instead of
the shift, it is convenient to parametrize a given gapped quantum Hall
state by a parameter $\kappa$ giving the offset between the total
charge $N_e$ and the total magnetic flux in the unit of flux quantum
$N_\phi$, on a sphere~\cite{Golkar:2014wwa}:
\begin{equation}\label{NNphik}
  N_e = \nu N_\phi + \kappa.
\end{equation}
To compare $\kappa$ with the shift of the corresponding QH state on
the LLL, one has to take into account two facts: there is a $\frac12$
offset between the definitions of the relativistic and nonrelativistic
filling factors, $\nu=\nu_{\rm NR}-\frac12$, and that Dirac particle
has a direct coupling to the spin connection.  As a result, the
connection between $\kappa$ and the shift $\mathcal S$
is~\cite{Golkar:2014wwa}
\begin{equation}\label{kappaS}
  \kappa = \nu_{\rm NR} (\mathcal S-1).
\end{equation}
In particular, if two states, ``1'' and ``2,'' are particle-hole
conjugates of each other, then $\nu^{\rm NR}_1 +\nu^{\rm NR}_2 =
  \nu^{\rm NR}_1 \mathcal S_1+\nu^{\rm NR}_2\mathcal S_2 =1$,
which in the relativistic notations become simply
\begin{equation}
  \nu_1+\nu_2 = \kappa_1+\kappa_2=0.
\end{equation}

Another quantity, which will not be discussed in this paper, is the
chiral central charge $c$.  The relationship between the relativistic
and nonrelativistic convention for this charge is $c = c_{\rm NR} -
\frac12$.  As for $\nu$ and $\kappa$, particle-hole conjugation flips
the sign of $c$.

\subsection{Discrete symmetries}

We recall the theory~(\ref{S}) has the following
symmetries~\cite{Deser:1981wh}:
\begin{itemize}
\item[(i)]
  Charge conjugation,
\begin{subequations}\label{C-AP}
\begin{align}
  \C A_\mu\C^{-1} &= -A_\mu\,, \\
  \C \Psi \C^{-1} &= \sigma^1 \Psi^*. \label{C-Psi}
\end{align}
\end{subequations}
\item[(ii)]
  Spatial parity $\x=(x,y)\to \x' = (x,-y)$,
\begin{subequations}\label{P-AP}
\begin{align}
  \P A_0(t,\x) \P^{-1} &= \phantom +  A_0(t,\x'),\\
  \P A_1(t,\x) \P^{-1} &= \phantom + A_1(t,\x'),\\
  \P A_2(t,\x) \P^{-1} &= -A_2(t,\x'),\\
  \P \Psi(t,\x) \P^{-1} &= \sigma^1\Psi(t,\x'). \label{P-Psi}
\end{align}
\end{subequations}
\item[(iii)]
  Time reversal $t\to -t$
\begin{subequations}\label{T-AP}
\begin{align}
  \T A_0(t,\x) \T^{-1} &= \phantom + A_0(-t,\x),\\
  \T A_i(t,\x) \T^{-1} &= - A_i(-t,\x),\\
  \T \Psi(t,\x)\T^{-1} &= -i\sigma^2 \Psi(-t,\x). \label{T-Psi}
\end{align}    
\end{subequations}
\end{itemize}
$\T$ is an anti-unitary operator ($\T i\T^{-1} =-i$).  These individual
symmetries are broken by the magnetic field $B$ which changes sign
under each of $\C$, $\P$, and $\T$.  However, $\C\P$, $\C\T$, and
$\P\T$ leave the magnetic field unchanged and hence remain the
discrete symmetries of the Dirac fermion in a magnetic
field~\cite{Golkar:2014wwa}.  We lose one symmetry because only two
out of these three are independent, e.g., $\C\T\sim(\C\P)(\P\T)$.  The
chemical potential further breaks $\C\P$ and $\C\T$, leaving $\P\T$ as
the only symmetry of the Dirac fermion in a magnetic field at nonzero
chemical potential.  As we are interested mostly in the charge
neutrality point, $\C\P$ and $\C\T$ will be exploited as
symmetries~\footnote{The parity anomaly~\cite{Niemi:1983rq,Redlich:1983dv}
  is not relevant for the current problem, as the gauge field propagates
  in (3+1) dimensions~\cite{Mulligan:2013he}.}.
  
When $e^2\ll1$, all interesting physics occurs on the $n=0$ Landau
level, and the $\C\P$, $\C\T$, and $\P\T$ operators map to symmetries of
the LLL Hamiltonian.  Let us choose the symmetric gauge, where the
single-particle orbitals on the LLL are $z^m e^{-|z|^2/4}$.  The field
$\Psi$ projected to the LLL can be written as
\begin{equation}\label{orbital}
  \Psi = \sum_m \begin{pmatrix}
    A_m z^m e^{-B|z|^2/4}
    \\ 0 \end{pmatrix} c_m\,,
\end{equation}
where $A_m$ is a normalization coefficient
and $c_m$ is the
operator annihilating a fermion on the orbital $m$. Then, from
Eqs.~(\ref{C-Psi}), (\ref{P-Psi}), and (\ref{T-Psi}), we find the
action of the symmetries on $c_m$,
\begin{align}
  \C\P c_m (\C\P)^{-1} &= c_m^\+\,,\\
  \P\T c_m (\P\T)^{-1} &= c_m\,, \\
  \C\T c_m (\C\T)^{-1} &= c_m^\+\,.
\end{align}
Note that $\C\T$ is different from $\C\P$: the former is an
anti-unitary transformation, while the latter is unitary.  In the
quantum Hall literature, by particle-hole symmetry, one normally has in
mind $\C\T$.  Likewise, $\P\T$ is not an identity operator: it replaces
the wave function of a given state, in the basis obtained by acting
$c_m^\dagger$'s on the vacuum, with its complex conjugate.  More precisely,
if $|f\>$ is the following quantum $N$-body state on the LLL,
\begin{equation}
  |f\> = \sum_{\{n_i\}} f_{n_1n_2\ldots n_N}
  c_{n_1}^\+ c_{n_2}^\+ \ldots
         c_{n_N}^\+ |0\>,
\end{equation}
then
\begin{equation}
  \P\T |f\> = \sum_{\{n_i\}} f_{n_1n_2\ldots n_N}^*
  c_{n_1}^\+ c_{n_2}^\+ \ldots
         c_{n_N}^\+ |0\>.
\end{equation}
Note that the Laughlin states~\cite{Laughlin:1983fy} and the
Moore-Read states~\cite{Moore:1991ks} are invariant under $\P\T$: in
both cases, the coefficients appearing in the holomorphic polynomial
in the wave function are all real.  We are unaware of any trial
wave function that is not invariant under $\P\T$.  Assuming $\P\T$
symmetry, $\C\P$ is equivalent to $\C\T$, thus, we will sometimes call
$\C\P$ the particle-hole symmetry, although in the literature the
latter usually corresponds to $\C\T$.

\subsection{Consequences of discrete symmetries for linear response}

Consider the linear response of a QH system to a small perturbation of
the electromagnetic potential: $j^\mu(q)=\Pi^{\mu\nu}(q)A_\nu(q)$,
$q_\mu\Pi^{\mu\nu}=0$.  Assuming rotational invariance, there are three
independent components of the polarization tensor $\Pi^{\mu\nu}$,
\begin{equation}\label{PiLRH}
  \Pi^{ij} = \frac{q^iq^j}{q^2}\Pi_{\rm L} +
  \left(\delta^{ij}-\frac{q^iq^j}{q^2} \right)\Pi_{\rm T}
  + \epsilon^{ij}\Pi_{\rm H} .
\end{equation}
The Hall conductivity $\sigma_{xy}$ is related to $\Pi_{\rm H}$ by
$\Pi_{\rm H}=\frac i{2\pi}\omega\sigma_{xy}$.  At nonzero chemical
potential, $\P\T$ is the only symmetry, and we get the constraints
$\Pi_{\rm L,T}(-\omega,q)=\Pi^*_{\rm L,T}(\omega,q)$ and
$\sigma_{xy}(-\omega,q) = \sigma_{xy}^*(\omega,q)$, but at charge
neutrality, $\C\P$ implies that the Hall conductivity vanishes,
\begin{equation}\label{sigmaxy0}
  \sigma_{xy}(\omega,q)=0,
\end{equation}
at any value of $\omega$ and $q$.  This is true not only for clean
systems but also in the presence of impurities, provided that the latter do not
statistically break the particle-hole symmetry.

For nonrelativistic fermions on the LLL, we can use
Eq.~(\ref{sxy-sNRxy}) to find, for $q=0$,
\begin{equation}
\sigma_{xy}^{\rm NR}(\omega)=\frac12\,,
\end{equation}
generalizing a result derived in Ref.~\cite{Kivelson:1997} to
nonzero frequencies.  For $q\neq0$, this relationship is replaced by a
linear constraint on $\sigma_{xy}$ and $\Pi_{\rm L}$~\cite{LevinSon}.

An explicitly particle-hole symmetric theory of the half-filled state
should imply Eq.~(\ref{sigmaxy0}) automatically.  We now construct
such a theory.


\section{Proposal for the low-energy effective field theory}
\label{sec:proposal}

Lacking a better way, we are going to simply guess the form of the
low-energy effective theory.  We start by stating a few requirements
that our theory should satisfy:
\begin{itemize}
\item[(i)] The theory should be invariant under $\C\P$ and $\P\T$.
\item[(ii)] At charge neutrality and nonzero magnetic field,
  the ground state should be a Fermi liquid.
\item[(iii)] The theory should explain the Jain sequences.
\end{itemize}
The requirement (ii) is particularly nontrivial, as Luttinger's
theorem requires the volume of the Fermi sphere to be proportional to
the density of some charge, but at charge neutrality, the
electromagnetic charge density is equal to zero.


It turns out that the three requirements above are satisfied by the
following action, which we propose as the low-energy effective theory:
\begin{multline}
  S_{\rm eff} =\int\!d^3x\left(
  i \bar\psi \gamma^\mu(\d_\mu + 2ia_\mu)\psi +
  \frac1{2\pi}\epsilon^{\mu\nu\lambda}A_\mu\d_\nu a_\lambda
  \right)\\
  - \frac1{4e^2}
  \int\!d^4x\, F_{\mu\nu}^2 + \cdots
  \label{L-a}
\end{multline}
Here, $\psi$ is a Dirac field describing the fermionic quasiparticle,
$a_\mu$ is an emergent gauge field, and $\ldots$ stands for other
terms, including a possible Maxwell kinetic term for $a_\mu$ and
interaction terms.  The quasiparticle $\psi$, which will also be
called the Dirac CF, has quantum numbers different from the 
electron: it is electrically neutral and carries
charge with respect to the emergent gauge field $a_\mu$.  Note that in
the HLR theory, the CF is also electrically neutral, a point which has
been emphasized by Read~\cite{Read-neutral1,Read-neutral2}.

Going from the microscopic action~(\ref{S}) to Eq.~(\ref{L-a}), it
looks as if the electron $\Psi$ has been stripped off of its charge,
whose dynamics is now governed by a dual gauge field $a_\mu$.
Interestingly, a similar phenomenon has recently been suggested to
occur in the ``composite Dirac liquid'' state~\cite{Mross:2014gla}---a
possible parent to several gapped topological phases of the strongly
interacting surface of 3D
TI \cite{Wang:2013uky,Bonderson:2013pla,Chen:2013jha,Metlitski:2013}.
As in our proposed effective theory,
the only low-energy mode of the composite Dirac liquid is a neutral
Dirac fermion.
While it is certain that the composite-fermion liquid phase cannot
occur in the theory~(\ref{S}) in the zero magnetic field in the limit
$e^2\ll1$, what we are suggesting is that as soon as one turns on a
magnetic field, the most convenient representation of the low-energy
dynamics is not in terms of the original fermion but in terms of a
composite Dirac fermion of the type considered in
Ref.~\cite{Mross:2014gla}.

In Eq.~(\ref{L-a}), we normalize the field $a_\mu$ so that a $2\pi$
magnetic flux of $a$ carries a unit electric charge [see
  Eq.~(\ref{j=da}) below].  With this normalization, the charge of the
$\psi$ field is fixed to 2.  As we will see below, this value of the
charge is required for the theory to be consistent with Jain's
sequences, the Fermi momentum of the half-filled state, and the $e/4$
charged excitation in the Pfaffian state.

\subsection{Discrete symmetries of the effective field theory}

It is easy to check that the Lagrangian~(\ref{L-a}) exhibits the full
set of discrete symmetries of the original theory~(\ref{S}), including
\begin{itemize}
\item[(i)] charge conjugation,
\begin{subequations}
\begin{align}
  \C a_\mu(t,\x) \C^{-1} &= - a_\mu(t,\x),\\
  \C\psi(t,\x)\C^{-1} &= \sigma^1\psi^*(t,\x);
\end{align}
\end{subequations}
\item[(ii)] spatial parity,
\begin{subequations}
\begin{align}
  \P a_0(t,\x) P^{-1} &= - a_0(t,\x'),\\
  \P a_1(t,\x) P^{-1} &= - a_1(t,\x'),\\
  \P a_2(t,\x) P^{-1} &= \phantom + a_2(t,\x'),\\
  \P \psi(t,\x)P^{-1} &= \psi^*(t,\x');
\end{align}
\end{subequations}
\item[(iii)] time reversal,
\begin{subequations}
\begin{align}
  \T a_0(t,\x) T^{-1} &=  - a_0(-t,\x),\\
  \T a_i(t,\x) T^{-1} & = \phantom + a_i(-t,\x), \\
  \T \psi(t,\x)T^{-1} &= \sigma^3\psi^*(-t,\x).
\end{align}
\end{subequations}
\end{itemize}
[We omit the transformation laws for $A_\mu$ which are the same as in
  Eqs.~(\ref{C-AP}), (\ref{P-AP}), and (\ref{T-AP})]. In particular,
our effective theory is invariant under $\C\P$ and $\P\T$.  We note
that the parity anomaly is avoided in this theory by the charge 2 of the
CF.
Note that the fermion mass term
$m\bar\psi\psi$ and the Chern-Simons term for $a_\mu$ are allowed by
$\P\T$ but forbidden by $\C\P$ or $\C\T$.  But we expect the terms
not forbidden by symmetries to appear in the ``$\cdots$'' in
Eq.~(\ref{L-a}).

It is instructive to write down the transformation law for $\psi$ under $\C\P$, $\C\T$,
\begin{subequations}
\begin{align}
  \C\P\psi(t,\x)(\C\P)^{-1} &= \sigma^1\psi(t,-\x),\\
  \C\T\psi(t,\x)(\C\T)^{-1} &= -i\sigma^2\psi(-t,\x),\\
  \P\T\psi(t,\x)(\P\T)^{-1} &= \sigma^3\psi(-t,-\x).
\end{align}
\end{subequations}
Note that under particle-hole symmetries ($\C\P$ and $\C\T$), $\psi$
does not transform into the complex conjugated field $\psi^*$ but
remains $\psi$.  In fact, the transformation laws of $\psi$ under
$\C\P$ and $\C\T$ are the same as those of the original electrons
$\Psi$ under $\P$ and $\T$ [Eqs.~(\ref{P-Psi}) and (\ref{T-Psi})].
Thus, particle-hole symmetry does not transform the
composite fermion $\psi$ into its antiparticle but leaves it as a
particle.  This result can also be seen from the fact that $a_0$, which is
proportional to the chemical potential of the CFs, does not change
sign under $\C\P$ and $\C\T$.

\subsection{Fermi liquid and Jain sequences}

To establish the nature of the QH state at charge neutrality, we first note
that the electromagnetic current defined by Eq.~(\ref{L-a}) is
\begin{equation}\label{j=da}
  j^\mu = \frac1{2\pi} \epsilon^{\mu\nu\lambda}\d_\nu a_\lambda\,.
\end{equation}
In particular, charge density is related to the emergent magnetic
field: $b=2\pi\rho$.  At charge neutrality, $\rho=0$, and therefore the
fermionic quasiparticles feel a zero magnetic field.

On the other hand, differentiating the action with respect to $a_0$,
one finds a relationship between the density of the Dirac CF and the
magnetic field,
\begin{equation}\label{tilden}
  \tilde\rho \equiv \< \bar\psi\gamma^0 \psi\> = \frac B{4\pi}\,.  
 \end{equation}
Therefore, in a nonzero magnetic field, the CFs have a finite density
and lives in a zero magnetic field.

Assuming no Cooper instability (a
possibility that we will consider later), the ground state is then a
Fermi liquid of the CFs, and Luttinger's theorem fixes the Fermi momentum to
the inverse of the magnetic length.  The value of the Fermi momentum is
the same as the value in the standard HLR theory~\cite{Halperin:1992mh}.
Note that the Fermi velocity $v_F$ of the CF
is, in general, different from 1; at weak coupling $v_F$, should scale
with the interaction energy: $v_F\sim e^2$.  The renormalization of
the Fermi velocity is related to Landau's Fermi liquid
parameters~\cite{Baym:1975va}.  The effective theory can be trusted to
describe the dynamics of the fermions near the Fermi surface but not
far away from it.  In principle, one should be able to reformulate
the theory completely in terms of the degree of freedom near the Fermi
surface~\cite{Polchinski:1992ed,Shankar:1993pf}, but since preserving
the information about the Berry phase (which will be important when we
discuss the Jain sequences) is quite nontrivial, we will not try to do
so.

The fermion quasiparticle is the $\C\P$ conjugate of itself.  To see
it, we note that $\C\P\psi(\C\P)=\sigma^1\psi$.  Consider a particle
moving with momentum $\p=(p,0)$, choosing the $\p$ to be invariant
under $y\to-y$.  Its wave function is an eigenvector of the Dirac
Hamiltonian $\bm{\sigma}\cdot{\bf p}=\sigma^1p$, which is $(1,0)^T$.
This wave function is invariant under $\C\P$.

Away from charge neutrality, the fermion quasiparticles feel an effective
magnetic field equal to
\begin{equation}\label{tildeB}
  \tilde B = - 2 \nabla\times a = -4\pi\rho.
\end{equation}
Equations~(\ref{tilden}) and (\ref{tildeB}) tell us that from the
point of view of the composite fermions, the notions of density and
magnetic field get swapped, similarly to what happens under
particle-vortex duality for bosons~\cite{FisherLee:1989}.
As a result, the filling factor of the
original electrons $\nu=2\pi\rho/B$ and the effective filling factor
of the fermion quasiparticles $\tilde\nu=2\pi\tilde\rho/\tilde B$ are
inversely proportional to each other,
  \begin{equation}
    2\nu = -\frac1{2\tilde\nu}\,.
\end{equation}
This means the Jain-sequence state with filling factor $\nu=\frac
n{2n+1}-\frac12=-\frac1{2(2n+1)}$ maps to an IQH state of the fermion
quasiparticle with $\tilde\nu = n+\frac12$.  Note that the filling
factor $\nu=\frac{n+1}{2n+1}-\frac12$ maps to
$\tilde\nu=-(n+\frac12)$, making explicit the fact that the two states
are particle-hole conjugates of each other.

\subsection{Shift of states on the Jain sequences.}

A nontrivial check for the proposed effective field theory is the
computation of the shift~\cite{Wen:1992ej} of the states on the Jain sequences
on a sphere.
Consider the $\nu_{\rm NR}=\frac{n+1}{2n+1}$ state.  The fermions $\psi$ are
effectively in the magnetic field of $2b$ in the IQH state with
$\nu_{\rm eff}=-(n+\frac12)$.  From solving the Dirac equation on a sphere,
the degeneracy of the $n$th Landau level is $N_\phi+2|n|$.  Therefore,
\begin{equation}
  \int\!d\x\, \< \psi^\+\psi\> = \left(n+\frac12\right) \int\!d\x\,
    \frac {2b}{2\pi} + n (n+1).
\end{equation}
But $\rho=b/2\pi$ and $\<\psi^\+\psi\>=B/4\pi$; therefore, we have
\begin{equation}
  \frac{N_\phi}2 = (2n+1) N_e + n(n+1).
\end{equation}
This means [see Eq.~(\ref{NNphik})]
\begin{equation}
  \nu = \frac1{2(2n+1)}\,, \qquad \kappa = - \frac{n(n+1)}{2n+1}\,,
\end{equation}
so $\nu_{\rm NR}=\frac{n+1}{2n+1}$ and, from Eq.~(\ref{kappaS}), $\mathcal S=-n+1$.  The shift
of the $\nu_{\rm NR}=\frac n{2n+1}$ can be computed similarly to be
$\mathcal S=n+2$.  These values of the shift coincide with what is
known about these states~\cite{Wen:1992ej}.

\subsection{Pfaffian, anti-Pfaffian, and a particle-hole symmetric non-Abelian
  state}

We can construct various gapped states by letting $\psi$
form a BCS pair.  Because the Dirac fermion has an additional Berry
phase around the origin in momentum space, pairing occurs in channels
with even angular momentum.

Consider first $s$-wave pairing, with the order parameter being
$\psi^T \sigma_2\psi$.  It is easy to see that the order parameter is
invariant under $\C\P$, $\C\T$, and $\P\T$ symmetries (combined with
phase rotations of $\psi$ when required).

Since this is a particle-hole symmetric state, it cannot be the
Pfaffian or anti-Pfaffian state.  In the conventional composite
fermion picture, this state can be understood as a state where the
composite fermions form Cooper pairs in the $p_x-ip_y$ channel with
the orbital angular momentum opposite to the Moore-Read state.  While
the conventional picture does not immediately tell us so, we find here
that this state coincides with its particle-hole conjugate.  In
particular, for this state, $\kappa=0$, corresponding to the shift
$\mathcal S=1$.  Due to the condensation of Cooper pairs that carry charge 4
under $a_\mu$, the vortices in the condensate correspond to
quasiparticles with charge $\pm1/4$.  The vortex has fermionic zero
mode and should have non-Abelian statistics.  This state is a
particle-hole symmetric Pfaffian-like state, which we will term the PH-Pfaffian.  In fact, its
construction parallels that of the time-reversal-invariant T-Pfaffian state on the
surface of TIs~\cite{Bonderson:2013pla,Chen:2013jha}.

The Pfaffian and anti-Pfaffian states correspond to pairing of $\psi$
in the $\ell=\pm2$ channels.  The vortices in these states also have
charges $\pm1/4$ as the $s$-wave paired state.  Further confirmations
are obtained when one computes the shift.  If one puts a $d$-wave
bosonic condensate on a sphere, without any magnetic field we would
have four vortices or antivortices.  Since each vortex has charge
$1/4$, this means $\kappa=\pm1$, which corresponds to $\mathcal S=3$
and $\mathcal S=-1$ [see Eq.~(\ref{kappaS})].  These are the values of
the shift for the Pfaffian and anti-Pfaffian states,
respectively~\footnote{If one would want to make the anti-Pfaffian
  state in the conventional CF theory, to match the shift, one would
  have to pair the CFs in an $f$-wave channel $\ell=-3$, compared to
  $\ell=+1$ for the Pfaffian state.}.

In real systems, for example, in the $\nu=\frac52$ QH state, whether
the $\ell=0$ or $\ell=\pm2$ is favored is the question about the
energy of the ground state and cannot be determined from general
principles.  It is possible that there is no pairing instability at
$\ell=0$, while there are instabilities at finite
$\ell$ (Ref.~\cite{Kohn:1965zz}).

\subsection{Electromagnetic response}

In the random phase approximation (RPA), one can relate the response
functions of electrons $\Pi_{\mu\nu}$ to the response functions of the
Dirac composite fermion, $\tilde\Pi_{\mu\nu}$.  We use the notation of
Eq.~(\ref{PiLRH}) and assume the Fermi velocity of the quasiparticles
to be much smaller than the speed of light, so the interaction is an
instantaneous interaction.  The RPA calculation can be done in a
straightforward way, and one finds the response functions of the
electrons to be
\begin{align}
  \Pi_{\rm L} &= -\frac{\omega^2}{16\pi^2\Delta} \tilde\Pi_{\rm L},
  \label{sxysbarxyL}\\
  \Pi_{\rm T} &= -\frac{\omega^2}{16\pi^2\Delta} \left(\tilde\Pi_{\rm T}
  -\frac{q^2V(q)}{16\pi^2} \right),
  \label{sxysbarxyT}\\
  \Pi_{\rm H} &= \phantom+ \frac{\omega^2}{16\pi^2\Delta}\tilde\Pi_{\rm H},
  \label{sxysbarxy}
\end{align}
where
\begin{equation}
  \Delta = \left(\tilde\Pi_{\rm T}- \frac{q^2V(q)}{16\pi^2}
  \right) \tilde\Pi_{\rm L} - |\Pi_{\rm H}|^2 .
\end{equation}
At $\nu=0$, for massless Dirac composite fermions, we have
$\tilde\Pi_{\rm H}=0$, and we find a vanishing Hall conductivity of the
Dirac electrons, $\Pi_{\rm H}=0$, as required by $\C\P$.  The formulas
for the other response functions simplify
\begin{align}
  \Pi_{\rm L} &= - \frac{\omega^2}{16\pi^2\tilde\Pi_{\rm T}-q^2V(q)}\,,\\
  \Pi_{\rm T} &= -\frac{\omega^2}{16\pi^2\tilde\Pi_{\rm L}}\,.
\end{align}
The first equation coincides with the formula in the theory proposed
in Ref.~\cite{Lee:1998zze}, but the second does not.  The
density-density correlation function $\Pi_{\rm L}$ shows the same
behavior in the infrared as in the HLR theory.  For example, because of
the Landau damping in $\tilde\Pi_{\rm T}$ (note that the CFs always
have a Fermi surface), $\Pi_{\rm L}$ has a pole at $\omega\sim-iq^2$
for the Coulomb interaction and behaves as $\Pi_{\rm L}\sim\omega q$ for
$\omega\gg q^2$, reproducing two key features of the HLR
theory~\cite{Halperin:1992mh}.

In the limit $q\to0$, the formulas simplify considerably.  Substituting
\begin{align}
  \Pi_{\rm T}& =\Pi_{\rm L} = \frac{i\omega}{2\pi}\sigma_{xx}\,,\\
  \Pi_{\rm H}&= \frac{i\omega}{2\pi}\sigma_{xy}\,,
\end{align}
where $\sigma_{xx}$ and $\sigma_{xy}$ are the longitudinal and Hall
conductivities for the electrons, in units of $e^2/h=1/2\pi$ (and
the version with a tilde for the composite fermions), we find [assuming
$V(q)$ diverges slower than $q^{-2}$ at $q\to0$]
\begin{align}
  \sigma_{xx} &= \frac14 \frac{\tilde\sigma_{xx}}{\tilde\sigma_{xx}^2
    + \tilde\sigma_{xy}^2}\,,\label{sxx_st}\\
  \sigma_{xy} &= -\frac14 \frac{\tilde\sigma_{xy}}{\tilde\sigma_{xx}^2
    + \tilde\sigma_{xy}^2} + \left[\tfrac12 \right]_{\rm NR}
  \label{sxy_st}\,.
\end{align}
The term $1/2$ on the right-hand side of Eq.~(\ref{sxy_st}) should be
included in the nonrelativistic case.  We concentrate on this case from now on. 
From Eqs.~(\ref{sxx_st}) and (\ref{sxy_st}), we can
express $\tilde\sigma_{xy}$ in terms of the measurable $\rho_{xx}$
and $\rho_{xy}$ as follows:
\begin{equation}\label{sst}
  \tilde\sigma_{xy} = \frac12 + \frac{\rho_{xy}-2}{\rho_{xx}^2+(\rho_{xy}-2)^2}
  \,.
\end{equation}
Typically, $\rho_{xx}$ is small, and in the regime $\rho_{xy}-2\sim
\rho_{xx}^2$ the above equation can be written as
\begin{equation}\label{D-vs-HLR}
  \frac{\rho_{xy}-2}{\rho_{xx}^2} = \tilde\sigma_{xy}-\frac12\,.
\end{equation}
Equation~(\ref{D-vs-HLR}) opens up a possibility to experimentally
distinguish the Dirac CFs and the standard HLR theory.  The HLR
theory, as we will argue in Sec.~\ref{sec:comp}, corresponds to
$\tilde\sigma_{xy}=1/2$; thus, the right-hand side of
Eq.~(\ref{D-vs-HLR}) is zero at exact half filling (e.g.,
$\rho_{xy}=2$ at exact half filling).  On the other hand, for a system
near particle-hole symmetry, $\tilde\sigma_{xy}$ is small and the
right-hand side should be close to $-1/2$.  One can thus plot the
left-hand side of Eq.~(\ref{D-vs-HLR}) as a function of the magnetic field
$B$ in a range of temperature where $\rho_{xx}$ depends nontrivially
on the temperature.  Both HLR and our theory then predict that all
curves go through one point, whose position on the horizontal axis is
the value of the magnetic field at exact half filling $B_{1/2}$, and the
position on the vertical axis is $0$ in the HLR theory and $-1/2$
in the theory of massless Dirac composite fermion.  When particle-hole
symmetry is not exact, the position of the point on the vertical axis
is $-\gamma/2\pi$ where $\gamma$ is the Berry phase of the
composite fermions around the Fermi disk (assuming Hall conductivity
is dominated by the Berry phase).  Furthermore, we expect the deviation
of $\gamma$ from $\pi$ to be proportional to the amount of
Landau-level mixing, i.e., the ratio of the Coulomb energy scale and the
cyclotron energy.  Note that, because of the smallness of $\rho_{xx}$,
distinguishing Berry phases of 0 and $\pi$ requires a sufficiently
accurate measurement of $\rho_{xy}$: for a relatively large
$\rho_{xx}\sim 0.1\,h/e^2$, an accuracy better than $5\times
10^{-3}\,h/e^2$ is needed.

The formulas~(\ref{sxx_st}) and (\ref{sxy_st}) are valid at zero
frequencies.  At finite frequencies, one needs to modify the theory for
it to be consistent with Galilean symmetry and Kohn's theorem.  The
modification and the results for the conductivities are discussed in
the Appendix.

\subsection{Infrared divergences}

One can directly check that despite the fact that the dynamical gauge
field $a_\mu$ does not have a Chern-Simons interaction, the propagator
of the transverse component of $a_i$ has the same infrared behavior as
in the HLR theory:
\begin{equation}
  \<a_\perp(0,-\q) a_\perp(0,\q)\> \sim \frac1{|\q|}\,.
\end{equation}
Therefore, we expect that our
theory would have the same logarithmic divergences as in the
HLR theory with Coulomb interaction.

\section{Connection with the fermionic Chern-Simons theory}
\label{sec:comp}

The QH physics of the Dirac fermion shares the same LLL limit with
nonrelativistic electrons.  The standard theory describing the FQH
states for nonrelativistic electrons is the fermionic CS (HLR) theory.
One can then ask the following question: can one connects the fermionic CS
theory and the relativistic theory~(\ref{L-a})?  This may seem
impossible since the former has a CS term in the Lagrangian, while the
latter does not.  However, we will argue that, it is at least possible
to go continuously from the relativistic theory to the conventional
nonrelativistic fermionic CS theory, if one allows breaking of
particle-hole symmetry.

The starting theory of the fermionic CS theory is the Lagrangian
\begin{equation}
  \mathcal L = i\psi^\+(\d_t-iA_0) \psi
  - \frac1{2m}|(\d_i-iA_i)\psi|^2  -\frac1{8\pi} AdA
  + \cdots .
\end{equation}
We have included a Chern-Simons term
$-\frac1{8\pi}AdA\equiv-\frac1{8\pi}\epsilon^{\mu\nu\lambda}A_\mu\d_\nu
A_\lambda$ to take into account the contribution of negative energy
states to the Hall conductivity.  We now use the standard flux
attachment procedure and attach two flux quanta to the fermion $\psi$.
The resulting action is then
\begin{multline}
  \mathcal L = i\psi^\+(\d_t-iA_0+ic_0) \psi
  - \frac1{2m}|(\d_i-iA_i+ic_i)\psi|^2 \\
  -\frac1{8\pi}AdA
  + \frac1{8\pi}cdc + \cdots,
\end{multline}
where $c_\mu$ is the statistical gauge field.
Now, changing notation,
\begin{equation}
  c_\mu = A_\mu + 2a_\mu\,, \\
\end{equation}
the Lagrangian becomes
\begin{multline}\label{S-CF}
  \mathcal L =  i\psi^\+(\d_t+2ia_0) \psi
  - \frac1{2m}|(\d_i+2ia_i)\psi|^2+ \frac1{2\pi} ada\\
  + \frac1{2\pi} Ada + \cdots. 
\end{multline}
The form of the action is very similar to Eq.~(\ref{L-a}), but there
are two crucial differences: (i) the CF is nonrelativistic, and (ii)
there is a Chern-Simons interaction $ada$.

We now suggest that theory~(\ref{S-CF}) can be obtained from
Eq.~(\ref{L-a}) by adding to the latter a $\C\P$-breaking mass term and
taking the large mass limit.  The mass term for the Dirac CF,
$-m\bar\psi\psi$, is generally expected once one relaxes the condition
of particle-hole symmetry (the coefficient of the Chern-Simons term
$ada$ cannot change continuously and thus has to remain zero).  Now
consider the regime where the mass $m$ is very large, and the CF is in
the nonrelativistic regime.  Let us remind ourselves that if one has a
massive Dirac field, coupled to a U(1) gauge field
\begin{equation}
  \mathcal L = i\bar\psi(\d_\mu - ia_\mu)\psi - m\bar\psi\psi
   +\mu\bar\psi\gamma^0\psi,
\end{equation}
and if the chemical potential is inside the gap, i.e., $|\mu|<|m|$, then
one can integrate out $\psi$ to obtain an effective action for
$a_\mu$~\cite{Niemi:1983rq,Redlich:1983dv},
\begin{equation}\label{induced-CS}
  - \frac1{8\pi} \frac{m}{|m|} ada.
\end{equation}
Now imagine that $\mu$ is slightly above the mass, $\mu>|m|$ but
$\mu-|m|\ll m$.  In this case, one has to take into account the gapless
fermions, which are nonrelativistic, but one should also not forget the
induced Chern-Simons term~(\ref{induced-CS}), which can be thought of
as coming from the filled Dirac sea and is largely unaffected by a
small density of added fermions.  Hence, low-energy dynamics is
described by
\begin{multline}
  \mathcal L = i\psi^\+ (\d_t-ia_0) \psi -\frac1{2|m|} |(\d_i-ia_i)^2\psi|^2\\
  - \frac1{8\pi} \frac{m}{|m|} ada + \cdots . 
\end{multline}
By the substitution $a\to-2a$, and assuming $m<0$, one obtains the
action~(\ref{S-CF}) before coupling to electromagnetism.  Thus, the
usual fermionic CS theory is obtained in a large mass, nonrelativistic
limit of the Dirac CF theory.  In other words, one can deform the
Dirac CF theory without a CS term into the conventional
nonrelativistic theory with a CS term, at the price of breaking
particle-hole symmetry.

This does not mean that we have been able to derive the effective
field theory~(\ref{L-a}) using the familiar flux-attachment procedure,
but now we can look at the old puzzle of Ref.~\cite{Kivelson:1997} in
a new light.  If one performs the RPA calculation using the
action~(\ref{S-CF}), we would find expressions identical to
Eqs.~(\ref{sxysbarxyL})--(\ref{sxysbarxy}) with the replacement
\begin{equation}
  \tilde \Pi_{\rm H} \to \frac{i\omega}{2\pi} \left( \sigma^{\rm CF}_{xy}
  + \frac 1 2 \right),
\end{equation}
where $\sigma_{xy}^{\rm CF}$ is the Hall conductivity of the CFs (as a
function of frequency and wave number) and $\frac12$ comes from the
Chern-Simons term $ada$.  To be consistent with Eq.~(\ref{sigmaxy0}),
$\sigma_{xy}^{\rm CF}$ must be equal to $-\frac12$ for all frequencies
and wave numbers.  However, at least in the regime $\omega\gg v_F q$,
the Hall conductivity of a Fermi liquid is expected to be zero.  This is
essentially the puzzle identified in Ref.~\cite{Kivelson:1997}, and
this puzzle would not exist if the low-energy dynamics is indeed
described by the Dirac composite fermion.
For the latter, the nonzero value of $\sigma_{x}^{\rm CF}$ can be
interpreted as coming from the Berry phase of the CF around the Fermi
disk.  It seems that the key to understanding the particle-hole
symmetry in the composite fermion theory is to understand the
appearance of this Berry phase.

\section{Conclusions}
\label{sec:concl}

We have presented evidence that the composite fermion of the
half-filled Landau level is a Dirac fermion, whose Dirac mass vanishes
when particle-hole symmetry is exact.  Although we did not try to
derive Eq.~(\ref{L-a}) microscopically, it is hoped that such a derivation
is possible, following one of the methods used in
Ref.~\cite{Mross:2014gla} for deriving the composite Dirac
liquid Hamiltonian.  We also notice that the LLL
Lagrangian~(\ref{L-LLL}) suggests that the Dirac spinor may involve, as
one of its components, the Lagrange multiplier $\chi$ that enforces the
LLL constraints.  For now, it is also clear that the conventional flux
attachment procedure cannot be applied to the quantum Hall effect for
a Dirac fermion in a particle-hole symmetric manner.

The proposed map between the electromagnetic current in the original
and low-energy effective theory,
$\bar\Psi\gamma^\mu\Psi=\frac1{2\pi}\epsilon^{\mu\nu\lambda}\d_\nu
a_\lambda$, is reminiscent of mirror
symmetry~\cite{Intriligator:1996ex} in three dimensions.  Recently,
Hook \emph{et al.}\ used mirror symmetry to argue that the ground state of
(2+1)-dimensional supersymmetric QED in the presence of magnetic impurities is
characterized by an emergent Fermi surface which encloses a volume
proportional to the magnetic field~\cite{Hook:2014dfa}.  The emergent
fermion is a fermionic vortex which was interpreted as a bound
state of a gaugino with a dual photon.  It is tempting to
speculate a connection between mirror symmetry and the physics of the
half-filled Landau level.


We have suggested that careful measurement of transport near half
filling can determine the Berry phase of the composite fermion around
the Fermi surface, and distinguishes the standard HLR scenario and the
scenario of Dirac CF.  Our theory implies the existence of a new
gapped phase, the PH-Pfaffian state, distinct from the Pfaffian and
anti-Pfaffian states by being its own particle-hole conjugate.  It
would be interesting
if there exist physical systems where such a phase is realized.
Besides GaAs and
surface of 3D TIs, systems with a tunable parameter like bilayer
graphene in a perpendicular magnetic
field~\cite{Tutuc_bilayer:2014,Kim_bilayer:2014} seem to be promising
venues to search for this new phase.

\acknowledgments

The author thanks Alexander Abanov, Clay C\'ordova, Eduardo Fradkin,
Siavash Golkar, Jainendra Jain, Woowon Kang, Michael Levin, Eun-Gook
Moon, Sergej Moroz, Sungjay Lee, Andrei Parnachev, Xiao-Liang Qi,
Subir Sachdev, Steve Simon, Boris Spivak, and Paul Wiegmann for
valuable discussions and comments.  This work is supported, in part,
by the US DOE Grant No.\ DE-FG02-13ER41958, MRSEC Grant No.\ DMR-1420709 by
the NSF, ARO MURI Grant No.\ 63834-PH-MUR, and by a Simons Investigator
grant from the Simons Foundation.

\emph{Note added:} After this work was completed, the author
became aware of Ref.~\cite{Barkeshli:2015afa}, which has some
overlap with the current paper.  In particular, the phase CFL$_2$ in
Fig.~5 of Ref.~\cite{Barkeshli:2015afa} differs from the picture
advocated here only by the presence of a second Fermi surface, coupled
to $a_\mu$ and carrying a trivial Berry phase.  The author thanks
Maissam Barkeshli and Michael Mulligan for illuminating discussions.
In addition, Ashwin Vishwanath informed the author that the
topological order of the PH-Pfaffian phase has been previously
propoosed within the context of topological
superconductors~\cite{Fidkowski:2013jua}.

\appendix

\section{Galilean invariance and Kohn's theorem}
\label{sec:Gal}

In the main text, we have avoided the issue of Galilean invariance and
the related Kohn's theorem.  Kohn's theorem states that in the absence
of disorder, the $q=0$, finite-frequency electromagnetic response of a
system of interacting nonrelativistic electrons is the same as of
noninteracting electrons.  In particular, the only pole in the
response is at the cyclotron frequency.  In the LLL limit $m\to0$,
the cyclotron frequency is infinite, and the response is especially
simple: $\sigma_{xx}(\omega)=0$, $\sigma_{xy}(\omega)=\nu$ (the
conductivities are measured in units of $e^2/h=1/2\pi$).  On the other
hand, Eq.~(\ref{sxx_st}) gives nonzero $\sigma_{xx}$ at finite
$\omega$.  The problem stems from the fact that the theory we were
using does not have the Galilean invariance of the original electrons.

The full solution to the problem of Galilean invariance, capable of
reproducing the more subtle effects such as the $q^2$ dependence of the
Hall conductivity of gapped states in the clean
limit~\cite{Hoyos:2011ez,Bradlyn:2012ea}, will be presented elsewhere.
Here, we only sketch a simplified version of the theory, adequate for
finding the $q=0$, finite-frequency response.  What follows is, in a
sense, a version (streamlined and trivially adapted to Dirac CFs) of
the phenomenological ``modified RPA'' (MRPA) proposal of Ref.~\cite{SH_MRPA}.
(At $q=0$, one does not need to be concerned with the issues that have
led to the ``MMRPA'' scheme of Ref.~\cite{MMRPA}.)

Galilean invariance is implemented by the introduction of a dynamic
field $v^i$, which transforms as a velocity under Galilean boosts.
The effective Lagrangian is schematically
\begin{equation}
  \mathcal L = \mathcal L (\psi,\psi^\+,a_\mu, v^i) + \frac1{2\pi}
  \epsilon^{\mu\nu\lambda}A_\mu \d_\nu a_\lambda\,.
\end{equation}
For modes near the Fermi surface, the coupling of $v^i$ to other
fields can be subsumed into a modified gauge potential,
\begin{align}
  \tilde a_0 &= a_0 - \frac14 m_* v^2\,,\\
  \tilde a_i &= a_i + \frac12 m_* v_i\,,
\end{align}
where $m_*$ is the effective mass of the composite fermions on the
Fermi surface.  In terms of $\tilde a_\mu$ the Lagrangian is
approximately
\begin{equation}
  \mathcal L = \mathcal L(\psi,\psi^\+,\tilde a_\mu) + \frac1{2\pi}
  \epsilon^{\mu\nu\lambda}A_\mu \d_\nu a_\lambda\,.
\end{equation}
where $\psi$ has a gauge coupling with $\tilde a_\mu$ with charge
$-2$.  (Note that the mixed Chern-Simons term still involves $a_\mu$,
not $\tilde a_\mu$.)  One can integrate out $v^i$ to obtain
interactions between the composite fermions, the major part of which
is a Landau-type interaction required to restore Galilean invariance.
To perform calculations in the RPA, it is, however, more convenient to
leave $v^i$ in the Lagrangian.

Differentiating the action with respect to $a_\mu$, we get the
relationship between the composite fermion density and the current with
the external field,
\begin{equation}
  \rho_{\rm CF} = \frac B{4\pi}\,,\qquad
  {\bf j}_{\rm CF} = \frac{{\bf E}\times{\bf \hat z}}{4\pi}\,.
\end{equation}
Differentiating with respect to $v^i$, one gets, on the other hand,
${\bf j}_{\rm CF} = \rho_{\rm CF} {\bf v}$, which implies
\begin{equation}
  {\bf v} = \frac{{\bf E}\times{\bf \hat z}}B\,.
\end{equation}
Denoting by $\tilde\sigma^{ij}$ the conductivity tensor of the CFs, we
have, for the composite fermion current,
\begin{equation}
  j^i_{\rm CF} = \frac{\tilde\sigma^{ij}}{2\pi} (-2\tilde e_j)
  =  \frac{\tilde\sigma^{ij}}{2\pi} (-2e_j + m_*\dot v_j).
\end{equation}
Furthermore, the physical electromagnetic current is related to the
field tensor constructed from $a_\mu$ [Eq.~(\ref{j=da})].  From the
formulas listed above, one can get the conductivities of the electrons,
\begin{subequations}\label{s_Gal}
\begin{align}
  \sigma_{xx} &= \frac14\frac{\tilde\sigma_{xx}}
        {\tilde\sigma_{xx}^2+\tilde\sigma_{xy}^2} + \frac{m_*}{2B}i\omega
        \label{sxx_Gal}\,,\\
  \sigma_{xy} &= -\frac14 \frac{\tilde\sigma_{xy}}{\tilde\sigma_{xx}^2
    + \tilde\sigma_{xy}^2} + \left[\tfrac12\right]_{\rm NR}\,.
  \label{sxy_Gal}
\end{align}
\end{subequations}
One sees that the only modification compared to Eqs.~(\ref{sxx_st})
and (\ref{sxy_st}) is the appearance of an additional term linear in
$\omega$ on the right-hand side of Eq.~(\ref{sxx_Gal}).

It is instructive to consider the Drude model for the composite
fermions, parametrizing disorders by a single relaxation time $\tau$.
In this approximation, the conductivities of the CFs are
\begin{align}
  \tilde\sigma_{xx} &= 2\pi\rho_{\rm CF}\,
  \frac{m_*(-i\omega+\tau^{-1})}{m_*^2(-i\omega+\tau^{-1})^2+4b^2}\,,\\
  \tilde\sigma_{xy} &= -2\pi\rho_{\rm CF}\,
  \frac{2b}{m_*^2(-i\omega+\tau^{-1})^2+4b^2}\,.
\end{align}
Substituting these expressions into Eqs.~(\ref{s_Gal}), we find
remarkably simple expressions,
\begin{align}
  \sigma_{xx} &= \frac{m_*}{2B\tau}\,,\\
  \sigma_{xy} &= \frac bB+\frac12 = \nu .
\end{align}
In the relaxation time approximation, the electron conductivities do
not depend on frequency, and the Hall conductivity is at its classical
value.  The clean case can be recovered by taking $\tau\to\infty$.
In this case, $\sigma_{xx}=0$ and $\sigma_{xy}=\nu$, as required by
Kohn's theorem at $m=0$.

\bibliography{dirac_cf-published}

\begin{thebibliography}{62}%
\makeatletter
\providecommand \@ifxundefined [1]{%
 \@ifx{#1\undefined}
}%
\providecommand \@ifnum [1]{%
 \ifnum #1\expandafter \@firstoftwo
 \else \expandafter \@secondoftwo
 \fi
}%
\providecommand \@ifx [1]{%
 \ifx #1\expandafter \@firstoftwo
 \else \expandafter \@secondoftwo
 \fi
}%
\providecommand \natexlab [1]{#1}%
\providecommand \enquote  [1]{``#1''}%
\providecommand \bibnamefont  [1]{#1}%
\providecommand \bibfnamefont [1]{#1}%
\providecommand \citenamefont [1]{#1}%
\providecommand \href@noop [0]{\@secondoftwo}%
\providecommand \href [0]{\begingroup \@sanitize@url \@href}%
\providecommand \@href[1]{\@@startlink{#1}\@@href}%
\providecommand \@@href[1]{\endgroup#1\@@endlink}%
\providecommand \@sanitize@url [0]{\catcode `\\12\catcode `\$12\catcode
  `\&12\catcode `\#12\catcode `\^12\catcode `\_12\catcode `\%12\relax}%
\providecommand \@@startlink[1]{}%
\providecommand \@@endlink[0]{}%
\providecommand \url  [0]{\begingroup\@sanitize@url \@url }%
\providecommand \@url [1]{\endgroup\@href {#1}{\urlprefix }}%
\providecommand \urlprefix  [0]{URL }%
\providecommand \Eprint [0]{\href }%
\providecommand \doibase [0]{http://dx.doi.org/}%
\providecommand \selectlanguage [0]{\@gobble}%
\providecommand \bibinfo  [0]{\@secondoftwo}%
\providecommand \bibfield  [0]{\@secondoftwo}%
\providecommand \translation [1]{[#1]}%
\providecommand \BibitemOpen [0]{}%
\providecommand \bibitemStop [0]{}%
\providecommand \bibitemNoStop [0]{.\EOS\space}%
\providecommand \EOS [0]{\spacefactor3000\relax}%
\providecommand \BibitemShut  [1]{\csname bibitem#1\endcsname}%
\let\auto@bib@innerbib\@empty
\bibitem [{\citenamefont {Tsui}\ \emph {et~al.}(1982)\citenamefont {Tsui},
  \citenamefont {Stormer},\ and\ \citenamefont {Gossard}}]{Tsui:1982yy}%
  \BibitemOpen
  \bibfield  {author} {\bibinfo {author} {\bibfnamefont {D.~C.}\ \bibnamefont
  {Tsui}}, \bibinfo {author} {\bibfnamefont {H.~L.}\ \bibnamefont {Stormer}}, \
  and\ \bibinfo {author} {\bibfnamefont {A.~C.}\ \bibnamefont {Gossard}},\
  }\bibfield  {title} {\enquote {\bibinfo {title} {{Two-Dimensional
  Magnetotransport in the Extreme Quantum Limit}},}\ }\href {\doibase
  10.1103/PhysRevLett.48.1559} {\bibfield  {journal} {\bibinfo  {journal}
  {Phys. Rev. Lett.}\ }\textbf {\bibinfo {volume} {48}},\ \bibinfo {pages}
  {1559} (\bibinfo {year} {1982})}\BibitemShut {NoStop}%
\bibitem [{\citenamefont {Laughlin}(1983)}]{Laughlin:1983fy}%
  \BibitemOpen
  \bibfield  {author} {\bibinfo {author} {\bibfnamefont {R.~B.}\ \bibnamefont
  {Laughlin}},\ }\bibfield  {title} {\enquote {\bibinfo {title} {{Anomalous
  Quantum Hall Effect: An Incompressible Quantum Fluid with Fractionally
  Charged Excitations}},}\ }\href {\doibase 10.1103/PhysRevLett.50.1395}
  {\bibfield  {journal} {\bibinfo  {journal} {Phys. Rev. Lett.}\ }\textbf
  {\bibinfo {volume} {50}},\ \bibinfo {pages} {1395} (\bibinfo {year}
  {1983})}\BibitemShut {NoStop}%
\bibitem [{\citenamefont {Jain}(1989)}]{Jain:1989tx}%
  \BibitemOpen
  \bibfield  {author} {\bibinfo {author} {\bibfnamefont {J.~K.}\ \bibnamefont
  {Jain}},\ }\bibfield  {title} {\enquote {\bibinfo {title} {{Composite-Fermion
  Approach for the Fractional Quantum Hall Effect}},}\ }\href {\doibase
  10.1103/PhysRevLett.63.199} {\bibfield  {journal} {\bibinfo  {journal} {Phys.
  Rev. Lett.}\ }\textbf {\bibinfo {volume} {63}},\ \bibinfo {pages} {199}
  (\bibinfo {year} {1989})}\BibitemShut {NoStop}%
\bibitem [{\citenamefont {Lopez}\ and\ \citenamefont
  {Fradkin}(1991)}]{Fradkin:1991wy}%
  \BibitemOpen
  \bibfield  {author} {\bibinfo {author} {\bibfnamefont {A.}~\bibnamefont
  {Lopez}}\ and\ \bibinfo {author} {\bibfnamefont {E.}~\bibnamefont
  {Fradkin}},\ }\bibfield  {title} {\enquote {\bibinfo {title} {{Fractional
  quantum Hall effect and Chern-Simons gauge theories}},}\ }\href {\doibase
  10.1103/PhysRevB.44.5246} {\bibfield  {journal} {\bibinfo  {journal} {Phys.
  Rev. B}\ }\textbf {\bibinfo {volume} {44}},\ \bibinfo {pages} {5246}
  (\bibinfo {year} {1991})}\BibitemShut {NoStop}%
\bibitem [{\citenamefont {Halperin}\ \emph {et~al.}(1993)\citenamefont
  {Halperin}, \citenamefont {Lee},\ and\ \citenamefont
  {Read}}]{Halperin:1992mh}%
  \BibitemOpen
  \bibfield  {author} {\bibinfo {author} {\bibfnamefont {B.~I.}\ \bibnamefont
  {Halperin}}, \bibinfo {author} {\bibfnamefont {P.~A.}\ \bibnamefont {Lee}}, \
  and\ \bibinfo {author} {\bibfnamefont {N.}~\bibnamefont {Read}},\ }\bibfield
  {title} {\enquote {\bibinfo {title} {{Theory of the half-filled Landau
  level}},}\ }\href {\doibase 10.1103/PhysRevB.47.7312} {\bibfield  {journal}
  {\bibinfo  {journal} {Phys. Rev. B}\ }\textbf {\bibinfo {volume} {47}},\
  \bibinfo {pages} {7312} (\bibinfo {year} {1993})}\BibitemShut {NoStop}%
\bibitem [{\citenamefont {Jain}(1997)}]{Jain-book}%
  \BibitemOpen
  \bibfield  {author} {\bibinfo {author} {\bibfnamefont {J.~K.}\ \bibnamefont
  {Jain}},\ }\href@noop {} {\emph {\bibinfo {title} {Composite Fermions}}}\
  (\bibinfo  {publisher} {Cambridge University Press},\ \bibinfo {address}
  {Cambridge, England},\ \bibinfo {year} {1997})\BibitemShut {NoStop}%
\bibitem [{\citenamefont {Willett}\ \emph {et~al.}(1990)\citenamefont
  {Willett}, \citenamefont {Paalanen}, \citenamefont {Ruel}, \citenamefont
  {West}, \citenamefont {Pfeiffer},\ and\ \citenamefont
  {Bishop}}]{Willett:1990}%
  \BibitemOpen
  \bibfield  {author} {\bibinfo {author} {\bibfnamefont {R.~L.}\ \bibnamefont
  {Willett}}, \bibinfo {author} {\bibfnamefont {M.~A.}\ \bibnamefont
  {Paalanen}}, \bibinfo {author} {\bibfnamefont {R.~R.}\ \bibnamefont {Ruel}},
  \bibinfo {author} {\bibfnamefont {K.~W.}\ \bibnamefont {West}}, \bibinfo
  {author} {\bibfnamefont {L.~N.}\ \bibnamefont {Pfeiffer}}, \ and\ \bibinfo
  {author} {\bibfnamefont {D.~J.}\ \bibnamefont {Bishop}},\ }\bibfield  {title}
  {\enquote {\bibinfo {title} {{Anomalous Sound Propagation at $\nu=\frac12$ in
  a 2D Electron Gas: Observation of a Spontaneously Broken Translational
  Symmetry?}}}\ }\href {\doibase 10.1103/PhysRevLett.65.112} {\bibfield
  {journal} {\bibinfo  {journal} {Phys. Rev. Lett.}\ }\textbf {\bibinfo
  {volume} {65}},\ \bibinfo {pages} {112} (\bibinfo {year} {1990})}\BibitemShut
  {NoStop}%
\bibitem [{\citenamefont {Kang}\ \emph {et~al.}(1993)\citenamefont {Kang},
  \citenamefont {Stormer}, \citenamefont {Pfeiffer}, \citenamefont {Baldwin},\
  and\ \citenamefont {West}}]{Kang:1993}%
  \BibitemOpen
  \bibfield  {author} {\bibinfo {author} {\bibfnamefont {W.}~\bibnamefont
  {Kang}}, \bibinfo {author} {\bibfnamefont {H.~L.}\ \bibnamefont {Stormer}},
  \bibinfo {author} {\bibfnamefont {L.~N.}\ \bibnamefont {Pfeiffer}}, \bibinfo
  {author} {\bibfnamefont {K.~W.}\ \bibnamefont {Baldwin}}, \ and\ \bibinfo
  {author} {\bibfnamefont {K.~W.}\ \bibnamefont {West}},\ }\bibfield  {title}
  {\enquote {\bibinfo {title} {{How Real Are Composite Fermions?}}}\ }\href
  {\doibase 10.1103/PhysRevLett.71.3850} {\bibfield  {journal} {\bibinfo
  {journal} {Phys. Rev. Lett.}\ }\textbf {\bibinfo {volume} {71}},\ \bibinfo
  {pages} {3850} (\bibinfo {year} {1993})}\BibitemShut {NoStop}%
\bibitem [{\citenamefont {Goldman}\ \emph {et~al.}(1994)\citenamefont
  {Goldman}, \citenamefont {Su},\ and\ \citenamefont {Jain}}]{Goldman:1994zz}%
  \BibitemOpen
  \bibfield  {author} {\bibinfo {author} {\bibfnamefont {V.~J.}\ \bibnamefont
  {Goldman}}, \bibinfo {author} {\bibfnamefont {B.}~\bibnamefont {Su}}, \ and\
  \bibinfo {author} {\bibfnamefont {J.~K.}\ \bibnamefont {Jain}},\ }\bibfield
  {title} {\enquote {\bibinfo {title} {{Detection of Composite Fermions by
  Magnetic Focusing}},}\ }\href {\doibase 10.1103/PhysRevLett.72.2065}
  {\bibfield  {journal} {\bibinfo  {journal} {Phys. Rev. Lett.}\ }\textbf
  {\bibinfo {volume} {72}},\ \bibinfo {pages} {2065} (\bibinfo {year}
  {1994})}\BibitemShut {NoStop}%
\bibitem [{\citenamefont {Moore}\ and\ \citenamefont
  {Read}(1991)}]{Moore:1991ks}%
  \BibitemOpen
  \bibfield  {author} {\bibinfo {author} {\bibfnamefont {G.~W.}\ \bibnamefont
  {Moore}}\ and\ \bibinfo {author} {\bibfnamefont {N.}~\bibnamefont {Read}},\
  }\bibfield  {title} {\enquote {\bibinfo {title} {{Nonabelions in the
  fractional quantum Hall effect}},}\ }\href {\doibase
  10.1016/0550-3213(91)90407-O} {\bibfield  {journal} {\bibinfo  {journal}
  {Nucl. Phys.}\ }\textbf {\bibinfo {volume} {B360}},\ \bibinfo {pages} {362}
  (\bibinfo {year} {1991})}\BibitemShut {NoStop}%
\bibitem [{\citenamefont {Read}\ and\ \citenamefont
  {Green}(2000)}]{Read:1999fn}%
  \BibitemOpen
  \bibfield  {author} {\bibinfo {author} {\bibfnamefont {N.}~\bibnamefont
  {Read}}\ and\ \bibinfo {author} {\bibfnamefont {D.}~\bibnamefont {Green}},\
  }\bibfield  {title} {\enquote {\bibinfo {title} {{Paired states of fermions
  in two dimensions with breaking of parity and time-reversal symmetries and
  the fractional quantum Hall effect}},}\ }\href {\doibase
  10.1103/PhysRevB.61.10267} {\bibfield  {journal} {\bibinfo  {journal} {Phys.
  Rev. B}\ }\textbf {\bibinfo {volume} {61}},\ \bibinfo {pages} {10267}
  (\bibinfo {year} {2000})},\ \Eprint {http://arxiv.org/abs/cond-mat/9906453}
  {cond-mat/9906453} \BibitemShut {NoStop}%
\bibitem [{\citenamefont {Girvin}(1984)}]{Girvin:1984zz}%
  \BibitemOpen
  \bibfield  {author} {\bibinfo {author} {\bibfnamefont {S.~M.}\ \bibnamefont
  {Girvin}},\ }\bibfield  {title} {\enquote {\bibinfo {title} {{Particle-hole
  symmetry in the anomalous quantum Hall effect}},}\ }\href {\doibase
  10.1103/PhysRevB.29.6012} {\bibfield  {journal} {\bibinfo  {journal} {Phys.
  Rev. B}\ }\textbf {\bibinfo {volume} {29}},\ \bibinfo {pages} {6012}
  (\bibinfo {year} {1984})}\BibitemShut {NoStop}%
\bibitem [{\citenamefont {Levin}\ \emph {et~al.}(2007)\citenamefont {Levin},
  \citenamefont {Halperin},\ and\ \citenamefont {Rosenow}}]{Levin:2007}%
  \BibitemOpen
  \bibfield  {author} {\bibinfo {author} {\bibfnamefont {M.}~\bibnamefont
  {Levin}}, \bibinfo {author} {\bibfnamefont {B.~I.}\ \bibnamefont {Halperin}},
  \ and\ \bibinfo {author} {\bibfnamefont {B.}~\bibnamefont {Rosenow}},\
  }\bibfield  {title} {\enquote {\bibinfo {title} {{Particle-Hole Symmetry and
  the Pfaffian State}},}\ }\href {\doibase 10.1103/PhysRevLett.99.236806}
  {\bibfield  {journal} {\bibinfo  {journal} {Phys. Rev. Lett.}\ }\textbf
  {\bibinfo {volume} {99}},\ \bibinfo {pages} {236806} (\bibinfo {year}
  {2007})},\ \Eprint {http://arxiv.org/abs/cond-mat/0707.0483}
  {cond-mat/0707.0483} \BibitemShut {NoStop}%
\bibitem [{\citenamefont {Lee}\ \emph {et~al.}(2007)\citenamefont {Lee},
  \citenamefont {Ryu}, \citenamefont {Nayak},\ and\ \citenamefont
  {Fisher}}]{SSLee:2007}%
  \BibitemOpen
  \bibfield  {author} {\bibinfo {author} {\bibfnamefont {S.~S.}\ \bibnamefont
  {Lee}}, \bibinfo {author} {\bibfnamefont {S.}~\bibnamefont {Ryu}}, \bibinfo
  {author} {\bibfnamefont {C.}~\bibnamefont {Nayak}}, \ and\ \bibinfo {author}
  {\bibfnamefont {M.~P.~A.}\ \bibnamefont {Fisher}},\ }\bibfield  {title}
  {\enquote {\bibinfo {title} {{Particle-Hole Symmetry and the $\nu=\frac52$
  Quantum Hall State}},}\ }\href {\doibase 10.1103/PhysRevLett.99.236807}
  {\bibfield  {journal} {\bibinfo  {journal} {Phys. Rev. Lett}\ }\textbf
  {\bibinfo {volume} {99}},\ \bibinfo {pages} {236807} (\bibinfo {year}
  {2007})},\ \Eprint {http://arxiv.org/abs/cond-mat/0707.0478}
  {cond-mat/0707.0478} \BibitemShut {NoStop}%
\bibitem [{\citenamefont {Kivelson}\ \emph {et~al.}(1997)\citenamefont
  {Kivelson}, \citenamefont {Lee}, \citenamefont {Krotov},\ and\ \citenamefont
  {Gan}}]{Kivelson:1997}%
  \BibitemOpen
  \bibfield  {author} {\bibinfo {author} {\bibfnamefont {S.~A.}\ \bibnamefont
  {Kivelson}}, \bibinfo {author} {\bibfnamefont {D.-H.}\ \bibnamefont {Lee}},
  \bibinfo {author} {\bibfnamefont {Y.}~\bibnamefont {Krotov}}, \ and\ \bibinfo
  {author} {\bibfnamefont {J.}~\bibnamefont {Gan}},\ }\bibfield  {title}
  {\enquote {\bibinfo {title} {{Composite-fermion Hall conductance at
  $\nu=\frac12$}},}\ }\href {\doibase 10.1103/PhysRevB.55.15552} {\bibfield
  {journal} {\bibinfo  {journal} {Phys. Rev. B}\ }\textbf {\bibinfo {volume}
  {55}},\ \bibinfo {pages} {15552} (\bibinfo {year} {1997})},\ \Eprint
  {http://arxiv.org/abs/cond-mat/9607153} {cond-mat/9607153} \BibitemShut
  {NoStop}%
\bibitem [{\citenamefont {Lee}(1998)}]{Lee:1998zze}%
  \BibitemOpen
  \bibfield  {author} {\bibinfo {author} {\bibfnamefont {D.-H.}\ \bibnamefont
  {Lee}},\ }\bibfield  {title} {\enquote {\bibinfo {title} {{Neutral Fermions
  at Filling Factor $\nu=1/2$}},}\ }\href {\doibase
  10.1103/PhysRevLett.80.4745} {\bibfield  {journal} {\bibinfo  {journal}
  {Phys. Rev. Lett.}\ }\textbf {\bibinfo {volume} {80}},\ \bibinfo {pages}
  {4745} (\bibinfo {year} {1998})},\ \Eprint
  {http://arxiv.org/abs/cond-mat/9709233} {cond-mat/9709233} \BibitemShut
  {NoStop}%
\bibitem [{\citenamefont {Rezayi}\ and\ \citenamefont
  {Haldane}(2000)}]{Rezayi:2000zz}%
  \BibitemOpen
  \bibfield  {author} {\bibinfo {author} {\bibfnamefont {E.~H.}\ \bibnamefont
  {Rezayi}}\ and\ \bibinfo {author} {\bibfnamefont {F.~D.~M.}\ \bibnamefont
  {Haldane}},\ }\bibfield  {title} {\enquote {\bibinfo {title} {{Incompressible
  Paired Hall State, Stripe Order, and the Composite Fermion Liquid Phase in
  Half-Filled Landau Levels}},}\ }\href {\doibase 10.1103/PhysRevLett.84.4685}
  {\bibfield  {journal} {\bibinfo  {journal} {Phys. Rev. Lett.}\ }\textbf
  {\bibinfo {volume} {84}},\ \bibinfo {pages} {4685} (\bibinfo {year}
  {2000})},\ \Eprint {http://arxiv.org/abs/cond-mat/9906137} {cond-mat/9906137}
  \BibitemShut {NoStop}%
\bibitem [{\citenamefont {Wu}\ \emph {et~al.}(1993)\citenamefont {Wu},
  \citenamefont {Dev},\ and\ \citenamefont {Jain}}]{Wu:1993}%
  \BibitemOpen
  \bibfield  {author} {\bibinfo {author} {\bibfnamefont {X.~G.}\ \bibnamefont
  {Wu}}, \bibinfo {author} {\bibfnamefont {G.}~\bibnamefont {Dev}}, \ and\
  \bibinfo {author} {\bibfnamefont {J.~K.}\ \bibnamefont {Jain}},\ }\bibfield
  {title} {\enquote {\bibinfo {title} {{Mixed-Spin Incompressible States in the
  Fractional Quantum Hall Effect}},}\ }\href {\doibase
  10.1103/PhysRevLett.71.153} {\bibfield  {journal} {\bibinfo  {journal} {Phys.
  Rev. Lett.}\ }\textbf {\bibinfo {volume} {71}},\ \bibinfo {pages} {153}
  (\bibinfo {year} {1993})}\BibitemShut {NoStop}%
\bibitem [{\citenamefont {Kamburov}\ \emph {et~al.}(2014)\citenamefont
  {Kamburov}, \citenamefont {Liu}, \citenamefont {Mueed}, \citenamefont
  {Shayegan}, \citenamefont {Pfeiffer}, \citenamefont {West},\ and\
  \citenamefont {Baldwin}}]{Baldwin:2014}%
  \BibitemOpen
  \bibfield  {author} {\bibinfo {author} {\bibfnamefont {D.}~\bibnamefont
  {Kamburov}}, \bibinfo {author} {\bibfnamefont {Y.}~\bibnamefont {Liu}},
  \bibinfo {author} {\bibfnamefont {M.~A.}\ \bibnamefont {Mueed}}, \bibinfo
  {author} {\bibfnamefont {M.}~\bibnamefont {Shayegan}}, \bibinfo {author}
  {\bibfnamefont {L.~N.}\ \bibnamefont {Pfeiffer}}, \bibinfo {author}
  {\bibfnamefont {K.~W.}\ \bibnamefont {West}}, \ and\ \bibinfo {author}
  {\bibfnamefont {K.~W.}\ \bibnamefont {Baldwin}},\ }\bibfield  {title}
  {\enquote {\bibinfo {title} {{What Determines the Fermi Wave Vector of
  Composite Fermions?}}}\ }\href {\doibase 10.1103/PhysRevLett.113.196801}
  {\bibfield  {journal} {\bibinfo  {journal} {Phys. Rev. Lett}\ }\textbf
  {\bibinfo {volume} {113}},\ \bibinfo {pages} {196801} (\bibinfo {year}
  {2014})},\ \Eprint {http://arxiv.org/abs/cond-mat/1406.2379}
  {cond-mat/1406.2379} \BibitemShut {NoStop}%
\bibitem [{\citenamefont {Novoselov}\ \emph {et~al.}(2005)\citenamefont
  {Novoselov}, \citenamefont {Geim}, \citenamefont {Morozov}, \citenamefont
  {Jiang}, \citenamefont {Katsnelson}, \citenamefont {Grigorieva},
  \citenamefont {Dubonos},\ and\ \citenamefont {Firsov}}]{Novoselov:2005kj}%
  \BibitemOpen
  \bibfield  {author} {\bibinfo {author} {\bibfnamefont {K.~S.}\ \bibnamefont
  {Novoselov}}, \bibinfo {author} {\bibfnamefont {A.~K.}\ \bibnamefont {Geim}},
  \bibinfo {author} {\bibfnamefont {S.~V.}\ \bibnamefont {Morozov}}, \bibinfo
  {author} {\bibfnamefont {D.}~\bibnamefont {Jiang}}, \bibinfo {author}
  {\bibfnamefont {M.~I.}\ \bibnamefont {Katsnelson}}, \bibinfo {author}
  {\bibfnamefont {I.~V}\ \bibnamefont {Grigorieva}}, \bibinfo {author}
  {\bibfnamefont {S.~V.}\ \bibnamefont {Dubonos}}, \ and\ \bibinfo {author}
  {\bibfnamefont {A.~A.}\ \bibnamefont {Firsov}},\ }\bibfield  {title}
  {\enquote {\bibinfo {title} {{Two-dimensional gas of massless Dirac fermions
  in graphene}},}\ }\href {\doibase 10.1038/nature04233} {\bibfield  {journal}
  {\bibinfo  {journal} {Nature}\ }\textbf {\bibinfo {volume} {438}},\ \bibinfo
  {pages} {197} (\bibinfo {year} {2005})},\ \Eprint
  {http://arxiv.org/abs/cond-mat/0509330} {cond-mat/0509330} \BibitemShut
  {NoStop}%
\bibitem [{\citenamefont {Zhang}\ \emph {et~al.}(2005)\citenamefont {Zhang},
  \citenamefont {Tan}, \citenamefont {Stormer},\ and\ \citenamefont
  {Kim}}]{Zhang:2005zz}%
  \BibitemOpen
  \bibfield  {author} {\bibinfo {author} {\bibfnamefont {Y.}~\bibnamefont
  {Zhang}}, \bibinfo {author} {\bibfnamefont {Y.-W.}\ \bibnamefont {Tan}},
  \bibinfo {author} {\bibfnamefont {H.~L.}\ \bibnamefont {Stormer}}, \ and\
  \bibinfo {author} {\bibfnamefont {P.}~\bibnamefont {Kim}},\ }\bibfield
  {title} {\enquote {\bibinfo {title} {{Experimental observation of the quantum
  Hall effect and and Berry's phase in graphene}},}\ }\href {\doibase
  10.1038/nature04235} {\bibfield  {journal} {\bibinfo  {journal} {Nature}\
  }\textbf {\bibinfo {volume} {438}},\ \bibinfo {pages} {201} (\bibinfo {year}
  {2005})},\ \Eprint {http://arxiv.org/abs/cond-mat/0509355} {cond-mat/0509355}
  \BibitemShut {NoStop}%
\bibitem [{\citenamefont {Du}\ \emph {et~al.}(2009)\citenamefont {Du},
  \citenamefont {Skachko}, \citenamefont {Duerr}, \citenamefont {Luican},\ and\
  \citenamefont {Andrei}}]{FQHE-graphene1}%
  \BibitemOpen
  \bibfield  {author} {\bibinfo {author} {\bibfnamefont {X.}~\bibnamefont
  {Du}}, \bibinfo {author} {\bibfnamefont {I.}~\bibnamefont {Skachko}},
  \bibinfo {author} {\bibfnamefont {F.}~\bibnamefont {Duerr}}, \bibinfo
  {author} {\bibfnamefont {A.}~\bibnamefont {Luican}}, \ and\ \bibinfo {author}
  {\bibfnamefont {E.~Y.}\ \bibnamefont {Andrei}},\ }\bibfield  {title}
  {\enquote {\bibinfo {title} {{Fractional quantum Hall effect and insulating
  phase of Dirac electrons in graphene}},}\ }\href {\doibase
  10.1038/nature08522} {\bibfield  {journal} {\bibinfo  {journal} {Nature}\
  }\textbf {\bibinfo {volume} {462}},\ \bibinfo {pages} {192} (\bibinfo {year}
  {2009})},\ \Eprint {http://arxiv.org/abs/0910.2532} {arXiv:0910.2532}
  \BibitemShut {NoStop}%
\bibitem [{\citenamefont {Bolotin}\ \emph {et~al.}(2009)\citenamefont
  {Bolotin}, \citenamefont {Ghahari}, \citenamefont {Shulman}, \citenamefont
  {Stormer},\ and\ \citenamefont {Kim}}]{FQHE-graphene2}%
  \BibitemOpen
  \bibfield  {author} {\bibinfo {author} {\bibfnamefont {K.~I.}\ \bibnamefont
  {Bolotin}}, \bibinfo {author} {\bibfnamefont {F.}~\bibnamefont {Ghahari}},
  \bibinfo {author} {\bibfnamefont {M.~D.}\ \bibnamefont {Shulman}}, \bibinfo
  {author} {\bibfnamefont {H.~L.}\ \bibnamefont {Stormer}}, \ and\ \bibinfo
  {author} {\bibfnamefont {P.}~\bibnamefont {Kim}},\ }\bibfield  {title}
  {\enquote {\bibinfo {title} {{Observation of the fractional quantum Hall
  effect in graphene}},}\ }\href {\doibase 10.1038/nature08582} {\bibfield
  {journal} {\bibinfo  {journal} {Nature}\ }\textbf {\bibinfo {volume} {462}},\
  \bibinfo {pages} {196} (\bibinfo {year} {2009})},\ \Eprint
  {http://arxiv.org/abs/0910.2763} {arXiv:0910.2763} \BibitemShut {NoStop}%
\bibitem [{\citenamefont {Analytis}\ \emph {et~al.}(2010)\citenamefont
  {Analytis}, \citenamefont {McDonald}, \citenamefont {Riggs}, \citenamefont
  {Chu}, \citenamefont {Boebinger},\ and\ \citenamefont
  {Fisher}}]{Boebinger:2010}%
  \BibitemOpen
  \bibfield  {author} {\bibinfo {author} {\bibfnamefont {J.~G.}\ \bibnamefont
  {Analytis}}, \bibinfo {author} {\bibfnamefont {R.~D.}\ \bibnamefont
  {McDonald}}, \bibinfo {author} {\bibfnamefont {S.~C.}\ \bibnamefont {Riggs}},
  \bibinfo {author} {\bibfnamefont {J.-H.}\ \bibnamefont {Chu}}, \bibinfo
  {author} {\bibfnamefont {G.~S.}\ \bibnamefont {Boebinger}}, \ and\ \bibinfo
  {author} {\bibfnamefont {I.~R.}\ \bibnamefont {Fisher}},\ }\bibfield  {title}
  {\enquote {\bibinfo {title} {{Two-dimensional surface state in the quantum
  limit of a topological insulator}},}\ }\href {\doibase 10.1038/nphys1861}
  {\bibfield  {journal} {\bibinfo  {journal} {Nature Phys.}\ }\textbf {\bibinfo
  {volume} {6}},\ \bibinfo {pages} {960} (\bibinfo {year} {2010})},\ \Eprint
  {http://arxiv.org/abs/1003.1713} {arXiv:1003.1713} \BibitemShut {NoStop}%
\bibitem [{\citenamefont {Khveshchenko}(2007)}]{Khveshchenko:2006}%
  \BibitemOpen
  \bibfield  {author} {\bibinfo {author} {\bibfnamefont {D.~V.}\ \bibnamefont
  {Khveshchenko}},\ }\bibfield  {title} {\enquote {\bibinfo {title} {{Composite
  Dirac fermions in graphene}},}\ }\href {\doibase 10.1103/PhysRevB.75.153405}
  {\bibfield  {journal} {\bibinfo  {journal} {Phys. Rev. B}\ }\textbf {\bibinfo
  {volume} {75}},\ \bibinfo {pages} {153405} (\bibinfo {year} {2007})},\
  \Eprint {http://arxiv.org/abs/cond-mat/0607174} {cond-mat/0607174}
  \BibitemShut {NoStop}%
\bibitem [{Note1()}]{Note1}%
  \BibitemOpen
  \bibinfo {note} {Note that in Read's LLL theory of the bosonic $\nu =1$
  state~\cite {Read:1998dn} there is also a gauge field with no Chern-Simons
  term.}\BibitemShut {Stop}%
\bibitem [{\citenamefont {Haldane}(2004)}]{Haldane:2004zz}%
  \BibitemOpen
  \bibfield  {author} {\bibinfo {author} {\bibfnamefont {F.~D.~M.}\
  \bibnamefont {Haldane}},\ }\bibfield  {title} {\enquote {\bibinfo {title}
  {{Berry Curvature on the Fermi Surface: Anomalous Hall Effect as a
  Topological Fermi-Liquid Property}},}\ }\href {\doibase
  10.1103/PhysRevLett.93.206602} {\bibfield  {journal} {\bibinfo  {journal}
  {Phys. Rev. Lett.}\ }\textbf {\bibinfo {volume} {93}},\ \bibinfo {pages}
  {206602} (\bibinfo {year} {2004})},\ \Eprint
  {http://arxiv.org/abs/cond-mat/0408417} {cond-mat/0408417} \BibitemShut
  {NoStop}%
\bibitem [{\citenamefont {Nguyen}\ \emph {et~al.}()\citenamefont {Nguyen},
  \citenamefont {Son},\ and\ \citenamefont {Wu}}]{Nguyen:2014uea}%
  \BibitemOpen
  \bibfield  {author} {\bibinfo {author} {\bibfnamefont {D.~X.}\ \bibnamefont
  {Nguyen}}, \bibinfo {author} {\bibfnamefont {D.~T.}\ \bibnamefont {Son}}, \
  and\ \bibinfo {author} {\bibfnamefont {C.}~\bibnamefont {Wu}},\ }\bibfield
  {title} {\enquote {\bibinfo {title} {{Lowest Landau Level Stress Tensor and
  Structure Factor of Trial Quantum Hall Wave Functions}},}\ }\href@noop {} {\
  }\Eprint {http://arxiv.org/abs/1411.3316} {arXiv:1411.3316} \BibitemShut
  {NoStop}%
\bibitem [{\citenamefont {Mart\'inez}\ and\ \citenamefont
  {Stone}(1993)}]{Martinez:1993xv}%
  \BibitemOpen
  \bibfield  {author} {\bibinfo {author} {\bibfnamefont {J.}~\bibnamefont
  {Mart\'inez}}\ and\ \bibinfo {author} {\bibfnamefont {M.}~\bibnamefont
  {Stone}},\ }\bibfield  {title} {\enquote {\bibinfo {title} {{Current
  operators in the lowest Landau level}},}\ }\href {\doibase
  10.1142/S0217979293003723} {\bibfield  {journal} {\bibinfo  {journal} {Int.
  J. Mod. Phys. B}\ }\textbf {\bibinfo {volume} {07}},\ \bibinfo {pages} {4389}
  (\bibinfo {year} {1993})}\BibitemShut {NoStop}%
\bibitem [{\citenamefont {Rajaraman}\ and\ \citenamefont
  {Sondhi}(1994)}]{RajaramanSondhi}%
  \BibitemOpen
  \bibfield  {author} {\bibinfo {author} {\bibfnamefont {R.}~\bibnamefont
  {Rajaraman}}\ and\ \bibinfo {author} {\bibfnamefont {S.~L.}\ \bibnamefont
  {Sondhi}},\ }\bibfield  {title} {\enquote {\bibinfo {title} {{Landau level
  mixing and solenoidal terms in lowest Landau level currents}},}\ }\href
  {\doibase 10.1142/S0217984994001072} {\bibfield  {journal} {\bibinfo
  {journal} {Mod. Phys. Lett. B}\ }\textbf {\bibinfo {volume} {08}},\ \bibinfo
  {pages} {1065} (\bibinfo {year} {1994})}\BibitemShut {NoStop}%
\bibitem [{\citenamefont {Golkar}\ \emph {et~al.}(2014)\citenamefont {Golkar},
  \citenamefont {Roberts},\ and\ \citenamefont {Son}}]{Golkar:2014wwa}%
  \BibitemOpen
  \bibfield  {author} {\bibinfo {author} {\bibfnamefont {S.}~\bibnamefont
  {Golkar}}, \bibinfo {author} {\bibfnamefont {M.~M.}\ \bibnamefont {Roberts}},
  \ and\ \bibinfo {author} {\bibfnamefont {D.~T.}\ \bibnamefont {Son}},\
  }\bibfield  {title} {\enquote {\bibinfo {title} {{Effective field theory of
  relativistic quantum Hall systems}},}\ }\href {\doibase
  10.1007/JHEP12(2014)138} {\bibfield  {journal} {\bibinfo  {journal} {J. High
  Energy Phys.}\ }\textbf {\bibinfo {volume} {12}},\ \bibinfo {pages} {138}
  (\bibinfo {year} {2014})},\ \Eprint {http://arxiv.org/abs/1403.4279}
  {arXiv:1403.4279} \BibitemShut {NoStop}%
\bibitem [{\citenamefont {Deser}\ \emph {et~al.}(1982)\citenamefont {Deser},
  \citenamefont {Jackiw},\ and\ \citenamefont {Templeton}}]{Deser:1981wh}%
  \BibitemOpen
  \bibfield  {author} {\bibinfo {author} {\bibfnamefont {S.}~\bibnamefont
  {Deser}}, \bibinfo {author} {\bibfnamefont {R.}~\bibnamefont {Jackiw}}, \
  and\ \bibinfo {author} {\bibfnamefont {S.}~\bibnamefont {Templeton}},\
  }\bibfield  {title} {\enquote {\bibinfo {title} {{Topologically Massive Gauge
  Theories}},}\ }\href {\doibase 10.1006/aphy.2000.6013,
  10.1016/0003-4916(82)90164-6} {\bibfield  {journal} {\bibinfo  {journal}
  {Ann. Phys. (N.Y.)}\ }\textbf {\bibinfo {volume} {140}},\ \bibinfo {pages}
  {372} (\bibinfo {year} {1982})}\BibitemShut {NoStop}%
\bibitem [{Note2()}]{Note2}%
  \BibitemOpen
  \bibinfo {note} {The parity anomaly~\cite {Niemi:1983rq,Redlich:1983dv} is
  not relevant for the current problem, as the gauge field propagates in (3+1)
  dimensions~\cite {Mulligan:2013he}.}\BibitemShut {Stop}%
\bibitem [{\citenamefont {Levin}\ and\ \citenamefont {Son}()}]{LevinSon}%
  \BibitemOpen
  \bibfield  {author} {\bibinfo {author} {\bibfnamefont {M.}~\bibnamefont
  {Levin}}\ and\ \bibinfo {author} {\bibfnamefont {D.~T.}\ \bibnamefont
  {Son}},\ }\href@noop {} {\bibinfo  {journal} {unpublished}\ }\BibitemShut
  {NoStop}%
\bibitem [{\citenamefont {Read}(1994)}]{Read-neutral1}%
  \BibitemOpen
\bibfield  {journal} {  }\bibfield  {author} {\bibinfo {author} {\bibfnamefont
  {N.}~\bibnamefont {Read}},\ }\bibfield  {title} {\enquote {\bibinfo {title}
  {{Theory of the half-filled Landau level}},}\ }\href {\doibase
  10.1088/0268-1242/9/11S/002} {\bibfield  {journal} {\bibinfo  {journal}
  {Semicond. Sci. Technol.}\ }\textbf {\bibinfo {volume} {9}},\ \bibinfo
  {pages} {1859} (\bibinfo {year} {1994})}\BibitemShut {NoStop}%
\bibitem [{\citenamefont {Read}(1996)}]{Read-neutral2}%
  \BibitemOpen
  \bibfield  {author} {\bibinfo {author} {\bibfnamefont {N.}~\bibnamefont
  {Read}},\ }\bibfield  {title} {\enquote {\bibinfo {title} {{Recent progress
  in the theory of composite fermions near even-denominator filling
  factors}},}\ }\href {\doibase 10.1016/0039-6028(96)00318-4} {\bibfield
  {journal} {\bibinfo  {journal} {Surface Sci.}\ }\textbf {\bibinfo {volume}
  {361/362}},\ \bibinfo {pages} {7} (\bibinfo {year} {1996})}\BibitemShut
  {NoStop}%
\bibitem [{\citenamefont {Mross}\ \emph {et~al.}(2015)\citenamefont {Mross},
  \citenamefont {Essin},\ and\ \citenamefont {Alicea}}]{Mross:2014gla}%
  \BibitemOpen
  \bibfield  {author} {\bibinfo {author} {\bibfnamefont {D.~F.}\ \bibnamefont
  {Mross}}, \bibinfo {author} {\bibfnamefont {A.}~\bibnamefont {Essin}}, \ and\
  \bibinfo {author} {\bibfnamefont {J.}~\bibnamefont {Alicea}},\ }\bibfield
  {title} {\enquote {\bibinfo {title} {{Composite Dirac Liquids: Parent States
  for Symmetric Surface Topological Order}},}\ }\href {\doibase
  10.1103/PhysRevX.5.011011} {\bibfield  {journal} {\bibinfo  {journal} {Phys.
  Rev. X}\ }\textbf {\bibinfo {volume} {5}},\ \bibinfo {pages} {011011}
  (\bibinfo {year} {2015})},\ \Eprint {http://arxiv.org/abs/1410.4201}
  {arXiv:1410.4201} \BibitemShut {NoStop}%
\bibitem [{\citenamefont {Wang}\ \emph {et~al.}(2013)\citenamefont {Wang},
  \citenamefont {Potter},\ and\ \citenamefont {Senthil}}]{Wang:2013uky}%
  \BibitemOpen
  \bibfield  {author} {\bibinfo {author} {\bibfnamefont {C.}~\bibnamefont
  {Wang}}, \bibinfo {author} {\bibfnamefont {A.~C.}\ \bibnamefont {Potter}}, \
  and\ \bibinfo {author} {\bibfnamefont {T.}~\bibnamefont {Senthil}},\
  }\bibfield  {title} {\enquote {\bibinfo {title} {{Gapped symmetry preserving
  surface state for the electron topological insulator}},}\ }\href {\doibase
  10.1103/PhysRevB.88.115137} {\bibfield  {journal} {\bibinfo  {journal} {Phys.
  Rev. B}\ }\textbf {\bibinfo {volume} {88}},\ \bibinfo {pages} {115137}
  (\bibinfo {year} {2013})},\ \Eprint {http://arxiv.org/abs/1306.3223}
  {arXiv:1306.3223} \BibitemShut {NoStop}%
\bibitem [{\citenamefont {Bonderson}\ \emph {et~al.}(2013)\citenamefont
  {Bonderson}, \citenamefont {Nayak},\ and\ \citenamefont
  {Qi}}]{Bonderson:2013pla}%
  \BibitemOpen
  \bibfield  {author} {\bibinfo {author} {\bibfnamefont {P.}~\bibnamefont
  {Bonderson}}, \bibinfo {author} {\bibfnamefont {C.}~\bibnamefont {Nayak}}, \
  and\ \bibinfo {author} {\bibfnamefont {X.-L.}\ \bibnamefont {Qi}},\
  }\bibfield  {title} {\enquote {\bibinfo {title} {{A time-reversal invariant
  topological phase at the surface of a 3D topological insulator}},}\ }\href
  {\doibase 10.1088/1742-5468/2013/09/P09016} {\bibfield  {journal} {\bibinfo
  {journal} {J. Stat. Mech.}\ }\textbf {\bibinfo {volume} {2013}},\ \bibinfo
  {pages} {P09016} (\bibinfo {year} {2013})},\ \Eprint
  {http://arxiv.org/abs/1306.3230} {arXiv:1306.3230} \BibitemShut {NoStop}%
\bibitem [{\citenamefont {Chen}\ \emph {et~al.}(2014)\citenamefont {Chen},
  \citenamefont {Fidkowski},\ and\ \citenamefont {Vishwanath}}]{Chen:2013jha}%
  \BibitemOpen
  \bibfield  {author} {\bibinfo {author} {\bibfnamefont {X.}~\bibnamefont
  {Chen}}, \bibinfo {author} {\bibfnamefont {L.}~\bibnamefont {Fidkowski}}, \
  and\ \bibinfo {author} {\bibfnamefont {A.}~\bibnamefont {Vishwanath}},\
  }\bibfield  {title} {\enquote {\bibinfo {title} {{Symmetry enforced
  non-Abelian topological order at the surface of a topological insulator}},}\
  }\href {\doibase 10.1103/PhysRevB.89.165132} {\bibfield  {journal} {\bibinfo
  {journal} {Phys. Rev. B}\ }\textbf {\bibinfo {volume} {89}},\ \bibinfo
  {pages} {165132} (\bibinfo {year} {2014})},\ \Eprint
  {http://arxiv.org/abs/1306.3250} {arXiv:1306.3250} \BibitemShut {NoStop}%
\bibitem [{\citenamefont {Metlitski}\ \emph {et~al.}()\citenamefont
  {Metlitski}, \citenamefont {Kane},\ and\ \citenamefont
  {Fisher}}]{Metlitski:2013}%
  \BibitemOpen
  \bibfield  {author} {\bibinfo {author} {\bibfnamefont {M.~A.}\ \bibnamefont
  {Metlitski}}, \bibinfo {author} {\bibfnamefont {C.~L.}\ \bibnamefont {Kane}},
  \ and\ \bibinfo {author} {\bibfnamefont {M.~P.~A.}\ \bibnamefont {Fisher}},\
  }\bibfield  {title} {\enquote {\bibinfo {title} {{A symmetry-respecting
  topologically-ordered surface phase of 3d electron topological
  insulators}},}\ }\href@noop {} {\ }\Eprint {http://arxiv.org/abs/1306.3286}
  {arXiv:1306.3286} \BibitemShut {NoStop}%
\bibitem [{\citenamefont {Baym}\ and\ \citenamefont
  {Chin}(1976)}]{Baym:1975va}%
  \BibitemOpen
  \bibfield  {author} {\bibinfo {author} {\bibfnamefont {G.}~\bibnamefont
  {Baym}}\ and\ \bibinfo {author} {\bibfnamefont {S.~A.}\ \bibnamefont
  {Chin}},\ }\bibfield  {title} {\enquote {\bibinfo {title} {{Landau Theory of
  Relativistic Fermi Liquids}},}\ }\href {\doibase
  10.1016/0375-9474(76)90513-3} {\bibfield  {journal} {\bibinfo  {journal}
  {Nucl. Phys.}\ }\textbf {\bibinfo {volume} {A262}},\ \bibinfo {pages} {527}
  (\bibinfo {year} {1976})}\BibitemShut {NoStop}%
\bibitem [{\citenamefont {Polchinski}()}]{Polchinski:1992ed}%
  \BibitemOpen
  \bibfield  {author} {\bibinfo {author} {\bibfnamefont {J.}~\bibnamefont
  {Polchinski}},\ }\bibfield  {title} {\enquote {\bibinfo {title} {{Effective
  field theory and the Fermi surface}},}\ }\href@noop {} {\ }\Eprint
  {http://arxiv.org/abs/hep-th/9210046} {hep-th/9210046} \BibitemShut {NoStop}%
\bibitem [{\citenamefont {Shankar}(1994)}]{Shankar:1993pf}%
  \BibitemOpen
  \bibfield  {author} {\bibinfo {author} {\bibfnamefont {R.}~\bibnamefont
  {Shankar}},\ }\bibfield  {title} {\enquote {\bibinfo {title}
  {{Renormalization group approach to interacting fermions}},}\ }\href
  {\doibase 10.1103/RevModPhys.66.129} {\bibfield  {journal} {\bibinfo
  {journal} {Rev. Mod. Phys.}\ }\textbf {\bibinfo {volume} {66}},\ \bibinfo
  {pages} {129} (\bibinfo {year} {1994})},\ \Eprint
  {http://arxiv.org/abs/cond-mat/9307009} {cond-mat/9307009} \BibitemShut
  {NoStop}%
\bibitem [{\citenamefont {Fisher}\ and\ \citenamefont
  {Lee}(1989)}]{FisherLee:1989}%
  \BibitemOpen
  \bibfield  {author} {\bibinfo {author} {\bibfnamefont {M.~P.~A.}\
  \bibnamefont {Fisher}}\ and\ \bibinfo {author} {\bibfnamefont {D.~H.}\
  \bibnamefont {Lee}},\ }\bibfield  {title} {\enquote {\bibinfo {title}
  {{Correspondence between two-dimensional bosons and a bulk superconductor in
  a magnetic field}},}\ }\href {\doibase 10.1103/PhysRevB.39.2756} {\bibfield
  {journal} {\bibinfo  {journal} {Phys. Rev. B}\ }\textbf {\bibinfo {volume}
  {39}},\ \bibinfo {pages} {2756} (\bibinfo {year} {1989})}\BibitemShut
  {NoStop}%
\bibitem [{\citenamefont {Wen}\ and\ \citenamefont {Zee}(1992)}]{Wen:1992ej}%
  \BibitemOpen
  \bibfield  {author} {\bibinfo {author} {\bibfnamefont {X.~G.}\ \bibnamefont
  {Wen}}\ and\ \bibinfo {author} {\bibfnamefont {A.}~\bibnamefont {Zee}},\
  }\bibfield  {title} {\enquote {\bibinfo {title} {{Shift and Spin Vector: New
  Topological Quantum Numbers for the Hall Fluids}},}\ }\href {\doibase
  10.1103/PhysRevLett.69.953} {\bibfield  {journal} {\bibinfo  {journal} {Phys.
  Rev. Lett.}\ }\textbf {\bibinfo {volume} {69}},\ \bibinfo {pages} {953}
  (\bibinfo {year} {1992})}\BibitemShut {NoStop}%
\bibitem [{Note3()}]{Note3}%
  \BibitemOpen
  \bibinfo {note} {If one would want to make the anti-Pfaffian state in the
  conventional CF theory, to match the shift, one would have to pair the CFs in
  an $f$-wave channel $\ell =-3$, compared to $\ell =+1$ for the Pfaffian
  state.}\BibitemShut {Stop}%
\bibitem [{\citenamefont {Kohn}\ and\ \citenamefont
  {Luttinger}(1965)}]{Kohn:1965zz}%
  \BibitemOpen
  \bibfield  {author} {\bibinfo {author} {\bibfnamefont {W.}~\bibnamefont
  {Kohn}}\ and\ \bibinfo {author} {\bibfnamefont {J.~M.}\ \bibnamefont
  {Luttinger}},\ }\bibfield  {title} {\enquote {\bibinfo {title} {{New
  Mechanism for Superconductivity}},}\ }\href {\doibase
  10.1103/PhysRevLett.15.524} {\bibfield  {journal} {\bibinfo  {journal} {Phys.
  Rev. Lett.}\ }\textbf {\bibinfo {volume} {15}},\ \bibinfo {pages} {524}
  (\bibinfo {year} {1965})}\BibitemShut {NoStop}%
\bibitem [{\citenamefont {Niemi}\ and\ \citenamefont
  {Semenoff}(1983)}]{Niemi:1983rq}%
  \BibitemOpen
  \bibfield  {author} {\bibinfo {author} {\bibfnamefont {A.~J.}\ \bibnamefont
  {Niemi}}\ and\ \bibinfo {author} {\bibfnamefont {G.~W.}\ \bibnamefont
  {Semenoff}},\ }\bibfield  {title} {\enquote {\bibinfo {title}
  {{Axial-Anomaly-Induced Fermion Fractionization and Effective Gauge-Theory
  Actions in Odd-Dimensional Space-Times}},}\ }\href {\doibase
  10.1103/PhysRevLett.51.2077} {\bibfield  {journal} {\bibinfo  {journal}
  {Phys. Rev. Lett.}\ }\textbf {\bibinfo {volume} {51}},\ \bibinfo {pages}
  {2077} (\bibinfo {year} {1983})}\BibitemShut {NoStop}%
\bibitem [{\citenamefont {Redlich}(1984)}]{Redlich:1983dv}%
  \BibitemOpen
  \bibfield  {author} {\bibinfo {author} {\bibfnamefont {A.~N.}\ \bibnamefont
  {Redlich}},\ }\bibfield  {title} {\enquote {\bibinfo {title} {{Parity
  violation and gauge noninvariance of the effective gauge field action in
  three dimensions}},}\ }\href {\doibase 10.1103/PhysRevD.29.2366} {\bibfield
  {journal} {\bibinfo  {journal} {Phys. Rev. D}\ }\textbf {\bibinfo {volume}
  {29}},\ \bibinfo {pages} {2366} (\bibinfo {year} {1984})}\BibitemShut
  {NoStop}%
\bibitem [{\citenamefont {Intriligator}\ and\ \citenamefont
  {Seiberg}(1996)}]{Intriligator:1996ex}%
  \BibitemOpen
  \bibfield  {author} {\bibinfo {author} {\bibfnamefont {K.~A.}\ \bibnamefont
  {Intriligator}}\ and\ \bibinfo {author} {\bibfnamefont {N.}~\bibnamefont
  {Seiberg}},\ }\bibfield  {title} {\enquote {\bibinfo {title} {{Mirror
  symmetry in three-dimensional gauge theories}},}\ }\href {\doibase
  10.1016/0370-2693(96)01088-X} {\bibfield  {journal} {\bibinfo  {journal}
  {Phys. Lett. B}\ }\textbf {\bibinfo {volume} {387}},\ \bibinfo {pages} {513}
  (\bibinfo {year} {1996})},\ \Eprint {http://arxiv.org/abs/hep-th/9607207}
  {hep-th/9607207} \BibitemShut {NoStop}%
\bibitem [{\citenamefont {Hook}\ \emph {et~al.}(2014)\citenamefont {Hook},
  \citenamefont {Kachru}, \citenamefont {Torroba},\ and\ \citenamefont
  {Wang}}]{Hook:2014dfa}%
  \BibitemOpen
  \bibfield  {author} {\bibinfo {author} {\bibfnamefont {A.}~\bibnamefont
  {Hook}}, \bibinfo {author} {\bibfnamefont {S.}~\bibnamefont {Kachru}},
  \bibinfo {author} {\bibfnamefont {G.}~\bibnamefont {Torroba}}, \ and\
  \bibinfo {author} {\bibfnamefont {H.}~\bibnamefont {Wang}},\ }\bibfield
  {title} {\enquote {\bibinfo {title} {{Emergent Fermi surfaces,
  fractionalization and duality in supersymmetric QED}},}\ }\href {\doibase
  10.1007/JHEP08(2014)031} {\bibfield  {journal} {\bibinfo  {journal} {J. High
  Energy Phys.}\ }\textbf {\bibinfo {volume} {08}},\ \bibinfo {pages} {031}
  (\bibinfo {year} {2014})},\ \Eprint {http://arxiv.org/abs/1401.1500}
  {arXiv:1401.1500} \BibitemShut {NoStop}%
\bibitem [{\citenamefont {Lee}\ \emph {et~al.}(2014)\citenamefont {Lee},
  \citenamefont {Fallahazad}, \citenamefont {Xue}, \citenamefont {Dillen},
  \citenamefont {Kim}, \citenamefont {Taniguchi}, \citenamefont {Watanabe},\
  and\ \citenamefont {Tutuc}}]{Tutuc_bilayer:2014}%
  \BibitemOpen
  \bibfield  {author} {\bibinfo {author} {\bibfnamefont {K.}~\bibnamefont
  {Lee}}, \bibinfo {author} {\bibfnamefont {B.}~\bibnamefont {Fallahazad}},
  \bibinfo {author} {\bibfnamefont {J.}~\bibnamefont {Xue}}, \bibinfo {author}
  {\bibfnamefont {D.~C.}\ \bibnamefont {Dillen}}, \bibinfo {author}
  {\bibfnamefont {K.}~\bibnamefont {Kim}}, \bibinfo {author} {\bibfnamefont
  {T.}~\bibnamefont {Taniguchi}}, \bibinfo {author} {\bibfnamefont
  {K.}~\bibnamefont {Watanabe}}, \ and\ \bibinfo {author} {\bibfnamefont
  {E.}~\bibnamefont {Tutuc}},\ }\bibfield  {title} {\enquote {\bibinfo {title}
  {{Chemical potential and quantum Hall ferromagnetism in bilayer graphene}},}\
  }\href {\doibase 10.1126/science.1251003} {\bibfield  {journal} {\bibinfo
  {journal} {Science}\ }\textbf {\bibinfo {volume} {345}},\ \bibinfo {pages}
  {58} (\bibinfo {year} {2014})},\ \Eprint {http://arxiv.org/abs/1401.0659}
  {arXiv:1401.0659} \BibitemShut {NoStop}%
\bibitem [{\citenamefont {Maher}\ \emph {et~al.}(2014)\citenamefont {Maher}
  \emph {et~al.}}]{Kim_bilayer:2014}%
  \BibitemOpen
  \bibfield  {author} {\bibinfo {author} {\bibfnamefont {P.}~\bibnamefont
  {Maher}} \emph {et~al.},\ }\bibfield  {title} {\enquote {\bibinfo {title}
  {{Tunable fractional quantum Hall phases in bilayer graphene}},}\ }\href
  {\doibase 10.1126/science.1252875} {\bibfield  {journal} {\bibinfo  {journal}
  {Science}\ }\textbf {\bibinfo {volume} {345}},\ \bibinfo {pages} {61}
  (\bibinfo {year} {2014})},\ \Eprint {http://arxiv.org/abs/1403.2112}
  {arXiv:1403.2112} \BibitemShut {NoStop}%
\bibitem [{\citenamefont {Barkeshli}\ \emph {et~al.}()\citenamefont
  {Barkeshli}, \citenamefont {Mulligan},\ and\ \citenamefont
  {Fisher}}]{Barkeshli:2015afa}%
  \BibitemOpen
  \bibfield  {author} {\bibinfo {author} {\bibfnamefont {M.}~\bibnamefont
  {Barkeshli}}, \bibinfo {author} {\bibfnamefont {M.}~\bibnamefont {Mulligan}},
  \ and\ \bibinfo {author} {\bibfnamefont {M.~P.~A.}\ \bibnamefont {Fisher}},\
  }\bibfield  {title} {\enquote {\bibinfo {title} {{Particle-Hole Symmetry and
  the Composite Fermi Liquid}},}\ }\href@noop {} {\ }\Eprint
  {http://arxiv.org/abs/1502.05404} {arXiv:1502.05404} \BibitemShut {NoStop}%
\bibitem [{\citenamefont {Fidkowski}\ \emph {et~al.}(2013)\citenamefont
  {Fidkowski}, \citenamefont {Chen},\ and\ \citenamefont
  {Vishwanath}}]{Fidkowski:2013jua}%
  \BibitemOpen
  \bibfield  {author} {\bibinfo {author} {\bibfnamefont {L.}~\bibnamefont
  {Fidkowski}}, \bibinfo {author} {\bibfnamefont {X.}~\bibnamefont {Chen}}, \
  and\ \bibinfo {author} {\bibfnamefont {A.}~\bibnamefont {Vishwanath}},\
  }\bibfield  {title} {\enquote {\bibinfo {title} {{Non-Abelian Topological
  Order on the Surface of a 3D Topological Superconductor from an Exactly
  Solved Model}},}\ }\href {\doibase 10.1103/PhysRevX.3.041016} {\bibfield
  {journal} {\bibinfo  {journal} {Phys. Rev. X}\ }\textbf {\bibinfo {volume}
  {3}},\ \bibinfo {pages} {041016} (\bibinfo {year} {2013})},\ \Eprint
  {http://arxiv.org/abs/1305.5851} {arXiv:1305.5851} \BibitemShut {NoStop}%
\bibitem [{\citenamefont {Hoyos}\ and\ \citenamefont
  {Son}(2012)}]{Hoyos:2011ez}%
  \BibitemOpen
  \bibfield  {author} {\bibinfo {author} {\bibfnamefont {C.}~\bibnamefont
  {Hoyos}}\ and\ \bibinfo {author} {\bibfnamefont {D.~T.}\ \bibnamefont
  {Son}},\ }\bibfield  {title} {\enquote {\bibinfo {title} {{Hall Viscosity and
  Electromagnetic Response}},}\ }\href {\doibase
  10.1103/PhysRevLett.108.066805} {\bibfield  {journal} {\bibinfo  {journal}
  {Phys. Rev. Lett.}\ }\textbf {\bibinfo {volume} {108}},\ \bibinfo {pages}
  {066805} (\bibinfo {year} {2012})},\ \Eprint {http://arxiv.org/abs/1109.2651}
  {arXiv:1109.2651} \BibitemShut {NoStop}%
\bibitem [{\citenamefont {Bradlyn}\ \emph {et~al.}(2012)\citenamefont
  {Bradlyn}, \citenamefont {Goldstein},\ and\ \citenamefont
  {Read}}]{Bradlyn:2012ea}%
  \BibitemOpen
  \bibfield  {author} {\bibinfo {author} {\bibfnamefont {B.}~\bibnamefont
  {Bradlyn}}, \bibinfo {author} {\bibfnamefont {M.}~\bibnamefont {Goldstein}},
  \ and\ \bibinfo {author} {\bibfnamefont {N.}~\bibnamefont {Read}},\
  }\bibfield  {title} {\enquote {\bibinfo {title} {{Kubo formulas for
  viscosity: Hall viscosity, Ward identities, and the relation with
  conductivity}},}\ }\href {\doibase 10.1103/PhysRevB.86.245309} {\bibfield
  {journal} {\bibinfo  {journal} {Phys. Rev. B}\ }\textbf {\bibinfo {volume}
  {86}},\ \bibinfo {pages} {245309} (\bibinfo {year} {2012})},\ \Eprint
  {http://arxiv.org/abs/1207.7021} {arXiv:1207.7021} \BibitemShut {NoStop}%
\bibitem [{\citenamefont {Simon}\ and\ \citenamefont
  {Halperin}(1993)}]{SH_MRPA}%
  \BibitemOpen
  \bibfield  {author} {\bibinfo {author} {\bibfnamefont {S.~H.}\ \bibnamefont
  {Simon}}\ and\ \bibinfo {author} {\bibfnamefont {B.~I.}\ \bibnamefont
  {Halperin}},\ }\bibfield  {title} {\enquote {\bibinfo {title}
  {{Finite-wave-vector electromagnetic response of fractional quantum Hall
  states}},}\ }\href {\doibase 10.1103/PhysRevB.48.17368} {\bibfield  {journal}
  {\bibinfo  {journal} {Phys. Rev. B}\ }\textbf {\bibinfo {volume} {48}},\
  \bibinfo {pages} {17368} (\bibinfo {year} {1993})},\ \Eprint
  {http://arxiv.org/abs/cond-mat/9307048} {cond-mat/9307048} \BibitemShut
  {NoStop}%
\bibitem [{\citenamefont {Simon}\ \emph {et~al.}(1996)\citenamefont {Simon},
  \citenamefont {Stern},\ and\ \citenamefont {Halperin}}]{MMRPA}%
  \BibitemOpen
  \bibfield  {author} {\bibinfo {author} {\bibfnamefont {S.~H.}\ \bibnamefont
  {Simon}}, \bibinfo {author} {\bibfnamefont {A.}~\bibnamefont {Stern}}, \ and\
  \bibinfo {author} {\bibfnamefont {B.~I.}\ \bibnamefont {Halperin}},\
  }\bibfield  {title} {\enquote {\bibinfo {title} {{Composite fermions with
  orbital magnetizations}},}\ }\href {\doibase 10.1103/PhysRevB.54.R11114}
  {\bibfield  {journal} {\bibinfo  {journal} {Phys. Rev. B}\ }\textbf {\bibinfo
  {volume} {54}},\ \bibinfo {pages} {R11114} (\bibinfo {year} {1996})},\
  \Eprint {http://arxiv.org/abs/cond-mat/9604103} {cond-mat/9604103}
  \BibitemShut {NoStop}%
\bibitem [{\citenamefont {Read}(1998)}]{Read:1998dn}%
  \BibitemOpen
  \bibfield  {author} {\bibinfo {author} {\bibfnamefont {N.}~\bibnamefont
  {Read}},\ }\bibfield  {title} {\enquote {\bibinfo {title}
  {{Lowest-Landau-level theory of the quantum Hall effect: The
  Fermi-liquid-like state of bosons at filling factor one}},}\ }\href {\doibase
  10.1103/PhysRevB.58.16262} {\bibfield  {journal} {\bibinfo  {journal} {Phys.
  Rev. B}\ }\textbf {\bibinfo {volume} {58}},\ \bibinfo {pages} {16262}
  (\bibinfo {year} {1998})},\ \Eprint {http://arxiv.org/abs/cond-mat/9804294}
  {cond-mat/9804294} \BibitemShut {NoStop}%
\bibitem [{\citenamefont {Mulligan}\ and\ \citenamefont
  {Burnell}(2013)}]{Mulligan:2013he}%
  \BibitemOpen
  \bibfield  {author} {\bibinfo {author} {\bibfnamefont {M.}~\bibnamefont
  {Mulligan}}\ and\ \bibinfo {author} {\bibfnamefont {F.~J.}\ \bibnamefont
  {Burnell}},\ }\bibfield  {title} {\enquote {\bibinfo {title} {{Topological
  insulators avoid the parity anomaly}},}\ }\href {\doibase
  10.1103/PhysRevB.88.085104} {\bibfield  {journal} {\bibinfo  {journal} {Phys.
  Rev. B}\ }\textbf {\bibinfo {volume} {88}},\ \bibinfo {pages} {085104}
  (\bibinfo {year} {2013})},\ \Eprint {http://arxiv.org/abs/1301.4230}
  {arXiv:1301.4230} \BibitemShut {NoStop}%
\end{thebibliography}%

\end{document}